\documentclass[10pt,twocolumn,showpacs,amsmath,amssymb,floatfix,superscriptaddress]{revtex4-2}

\usepackage{graphicx}% Include figure files
\usepackage{float}
\usepackage{dcolumn}% Align table columns on decimal point
\usepackage{bm}
\usepackage{color} 
\usepackage{txfonts}
\usepackage{microtype}
\usepackage{hyperref}
\usepackage[english]{babel}
\usepackage{slashed}
\usepackage{gensymb}
\usepackage{epsfig}
\usepackage[normalem]{ulem}
\usepackage{amsmath}
\usepackage{array}
\usepackage{booktabs}
\usepackage[english]{babel}
\usepackage{multirow}
\usepackage{CJKutf8}
\allowdisplaybreaks[4]

\begin{document}
%\begin{CJK}{UTF8}{gbsn}	
\renewcommand{\figurename}{FIG}	

\title{\emph{In-situ} Generation of Polarized Attosecond Positron Bunches via Radiation-Wakefield Induced Two-Photon Pairs}% Force line breaks with \\

\author{Zhen-Ke Dou}
\affiliation{Ministry of Education Key Laboratory for Nonequilibrium Synthesis and Modulation of Condensed Matter, State key laboratory of electrical insulation and power equipment, Shaanxi Province Key Laboratory of Quantum Information and Quantum Optoelectronic Devices, School of Physics, Xi'an Jiaotong University, Xi'an 710049, China}

\author{Qian Zhao}\email{zhaoq2019@xjtu.edu.cn}
\affiliation{Ministry of Education Key Laboratory for Nonequilibrium Synthesis and Modulation of Condensed Matter, State key laboratory of electrical insulation and power equipment, Shaanxi Province Key Laboratory of Quantum Information and Quantum Optoelectronic Devices, School of Physics, Xi'an Jiaotong University, Xi'an 710049, China}

\author{Hao-Tian Wang}
\affiliation{Ministry of Education Key Laboratory for Nonequilibrium Synthesis and Modulation of Condensed Matter, State key laboratory of electrical insulation and power equipment, Shaanxi Province Key Laboratory of Quantum Information and Quantum Optoelectronic Devices, School of Physics, Xi'an Jiaotong University, Xi'an 710049, China}

\author{Feng Wan}
\affiliation{Ministry of Education Key Laboratory for Nonequilibrium Synthesis and Modulation of Condensed Matter, State key laboratory of electrical insulation and power equipment, Shaanxi Province Key Laboratory of Quantum Information and Quantum Optoelectronic Devices, School of Physics, Xi'an Jiaotong University, Xi'an 710049, China}

\author{Jian-Xing Li}\email{jianxing@xjtu.edu.cn}
\affiliation{Ministry of Education Key Laboratory for Nonequilibrium Synthesis and Modulation of Condensed Matter, State key laboratory of electrical insulation and power equipment, Shaanxi Province Key Laboratory of Quantum Information and Quantum Optoelectronic Devices, School of Physics, Xi'an Jiaotong University, Xi'an 710049, China}	
\affiliation{Department of Nuclear Physics, China Institute of Atomic Energy, P. O. Box 275(7), Beijing 102413, China}

%\date{\today}% It is always \today, today, but any date may be explicitly specified
	
\begin{abstract}

High-energy, dense, spin-polarized positron beams with attosecond duration are highly desirable for advanced accelerator physics, laboratory astrophysics, and ultrafast matter–antimatter studies, yet remain beyond the capability of conventional source and injection-based plasma-acceleration schemes. We propose an integrated \emph{in-situ} method for production, trapping, and accelerating polarized attosecond positron bunches in a single plasma stage. An ultraintense hollow laser propagating in a plasma channel drives a radiative wakefield, where longitudinally injected electrons emit a dense MeV $\gamma$-ray bath through nonlinear Compton scattering. Subsequent photon–photon collisions induce linear Breit–Wheeler pair production, while linear Compton scattering and nonlinear radiative processes mediate polarization transfer to the produced positrons. Spin-resolved QED particle-in-cell simulations indicate that the laser–modulated wakefield naturally captures positrons born in the frontier bubble wake, forming collimated bunch trains with hundred-attosecond duration, GeV-level energies, and densities up to $10^{17}\,\mathrm{cm^{-3}}$. Parameter scans demonstrate robustness against variations of laser intensity and plasma-channel density. This single-shot, source-free approach simultaneously addresses positron production, polarization, temporal compression, and plasma injection, offering a compact route toward polarized attosecond antimatter beams.

\end{abstract}
	
\maketitle

\section{Introduction}
Relativistic positron beams underpin a broad spectrum of frontier research, from precision electroweak physics at future $e^+e^-$ colliders to positron annihilation spectroscopy of defects and nanostructures, positron emission tomography, and trapped-antimatter tests of \emph{CPT} symmetry and the weak equivalence principle \cite{danielson2015Plasma,an2019Precision,fajans2020Plasma,chaikovska2022Positron,cao2024Positron,si2024Research}. They are also becoming indispensable tools for laboratory studies of electron--positron pair plasmas, which are central to the physics of gamma-ray bursts, pulsar magnetospheres, and relativistic jets \cite{ruffini2010electron,kumar2015physics,sarri2015Generation,stoneking2020new,hada2024M87,arrowsmith2024laboratory}. Beyond charge and energy, two additional phase-space dimensions have emerged as decisive: spin polarization and ultrashort temporal structure. Polarized positron beams with polarization fractions exceeding $\sim 30\%$ and densities above $10^{16}~\mathrm{cm}^{-3}$ are required for hadronic structure measurements, chiral tests of the Standard Model, and polarized muon production \cite{abbott2016Production,boscolo2020Muon,dou2025Compact}. Concurrently, femtosecond-to-attosecond positron bunches would enable ultrafast pump--probe measurements of matter--antimatter dynamics and naturally match the $10$--$100~\mathrm{fs}$ periods characteristic of plasma wakefields \cite{alejo2019Laser,streeter2024Narrow}.

Conventional positron sources, such as accelerator-based pair production and radioactive decay, typically provide fluxes of only \(10^6\)--\(10^8~\mathrm{e^+/s}\), largely limited by low trapping and collection efficiencies, and remain difficult to tailor into ultrabright femtosecond positron beams \cite{danielson2015Plasma,fajans2020Plasma,hessami2023Compact}. High-power lasers have opened a complementary route to compact relativistic positron sources. In the experimentally established Bethe--Heitler (BH) scheme, laser-accelerated electrons generate bremsstrahlung photons in high-\(Z\) targets, which subsequently decay into \(e^+e^-\) pairs \cite{sarri2013TableTop,chen2015Scaling,wu2016Optimization,chen2023Perspectives}. These laser-driven BH sources have enabled high-density  positron production, yet the resulting beams usually possess broad energy spectra, relativistic effective temperatures of \(1\)--\(20~\mathrm{MeV}\), and large divergences of \(10^\circ\)--\(40^\circ\), making beam capture and phase-space control highly challenging \cite{alejo2019Laser,streeter2024Narrow}.

With the advent of multi-PW laser facilities capable of reaching intensities above \(10^{23}\,\mathrm{W\,cm^{-2}}\) \cite{danson2019Petawatt,yoon2021Realization,yamanouchi2021Progress,radier2022Peak}, laser--matter interactions are entering the strong-field QED regime. In this regime, nonlinear Compton scattering (NCS) and nonlinear Breit--Wheeler (NBW) pair production can efficiently convert laser energy into ultrabright \(\gamma\)-rays and dense relativistic \(e^+e^-\) pairs \cite{sun2022Production,fedotov2023Advances,yu2024Bright}. Laser-irradiated solids and near-critical-density plasmas further enhance pair yields through radiation-reaction trapping, prolific photon emission, and QED cascades \cite{ridgers2012Dense,zhu2016Dense}. More recently, polarization-resolved QED-PIC simulations have shown that such positrons may acquire sizable spin polarization through spin-dependent pair creation and radiative spin dynamics in ultraintense laser--plasma interactions \cite{song2022Dense,xue2023Generation}. Below the efficient NBW threshold, dense counterpropagating \(\gamma\)-ray populations generated by \(10^{22}\,\mathrm{W\,cm^{-2}}\)-class lasers can also trigger linear Breit--Wheeler (LBW) pair creation, which may dominate over NBW and BH channels in structured or near-critical plasmas \cite{he2021Dominance,sugimoto2023Positron}; laser-solid studies indicate a transition from LBW- to NBW-dominated production at \(a_0\sim400\)--500 \cite{song2024From}. Together with polarization-resolved linear-QED cascades, this provides a promising path toward polarized positron sources on current 10-PW platforms \cite{zhao2023Cascade,dou2026Generation}.

Plasma wakefield acceleration (PWFA) offers a promising route for further accelerating laser-driven positron sources, owing to its ultrahigh accelerating gradients exceeding $100~\mathrm{GV/m}$ and intrinsically femtosecond-scale acceleration structure \cite{cao2024Positron}. However, laser-driven positron sources referred above generally produce highly divergent, broadband, and relativistic pair beams. Extracting a clean positron bunch from such mixed $e^+e^-$ plasmas using magnetic spectrometers, and subsequently synchronizing, focusing, and injecting it into a PWFA, remain major challenges \cite{alejo2019Laser,streeter2024Narrow}. Current PWFA-based positron acceleration studies mainly rely on either externally pre-accelerated GeV-class positron beams, which can excite self-loaded wakefields capable of focusing and accelerating the beam itself \cite{corde2015Multi,doche2017Acceleration,silva2023Positron}, or specially tailored plasma structures, such as hollow plasma channels that provide accelerating fields with vanishing transverse forces \cite{gessner2016Demonstration,silva2021Stable,zhou2021High}. Alternatively, hollow-beam drivers can excite doughnut-shaped wakefields with simultaneous accelerating and focusing phases for positrons near the front of the bubble wake \cite{vieira2014Nonlinear,jain2015Positron,liu2022Trapping,sun2025Generation}. Despite these advances, all these schemes require a preformed, phase-space-controlled positron bunch and stringent spatiotemporal matching to the wakefield, which is particularly difficult for laser-driven positron beams. These limitations motivate an integrated, single-stage scheme in which positron production, capture, and acceleration are combined within the same laser-plasma interaction.

% All-optical schemes combining Bethe--Heitler production and direct laser or plasma acceleration have also been proposed \cite{xu2020New,martinez2023Creation,terzani2023Compact,martinez2025Direct}, and laser-wakefield-driven positron sources have recently produced narrow-band, low-emittance, femtosecond positron beamlets near $600~\mathrm{MeV}$ \cite{streeter2024Narrow,alejo2019Laser}.

%These advances suggest a compact route toward high-brightness polarized positron sources for collider physics and laboratory astrophysics.

\begin{figure}[t!] 
	\begin{center}
		\includegraphics[width=\linewidth]{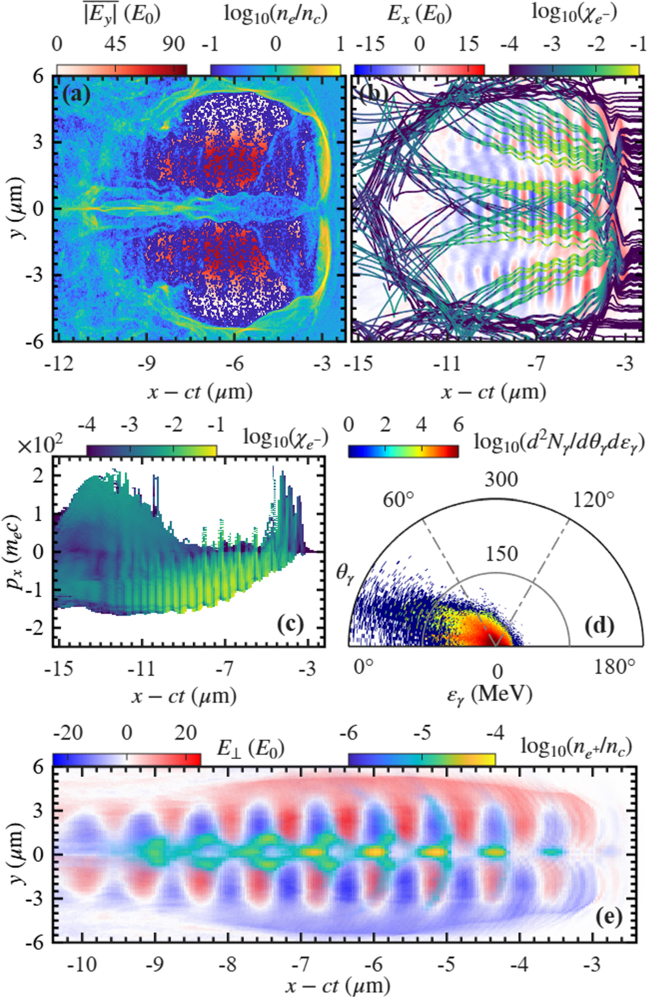}
		\caption{(a) Snapshots of the laser field $\overline{|E_y|}$ overlaid with plasma electron density $n_e$ at $t=40\:T_0$. (b) Corresponding longitudinal field $E_x$ at $t=40\:T_0$. The trajectories represent randomly selected plasma electrons, with a subset being accelerated and injected into the bubble. The color scale represents their quantum parameter $\chi_{e^-}$. (c) The distribution of the quantum parameter $\chi_{e^-}$ for electrons injected into the bubble. (d) Density distributions of gamma photons in the $\varepsilon_\gamma-\theta_\gamma$ plane of spherical coordinates at $t=40\:T_0$. (e) Transverse field $E_\perp=E_y-cB_z$ at $t=40\:T_0$, overlaid with the density of accelerated and focused positron bunch.}\label{fig1}
	\end{center}
\end{figure}
In this paper, we proposed a method for \emph{in-situ} generation of polarized attosecond positron bunches within a single laser-plasma stage, thereby eliminating the need for an external positron source and the associated injection complexity. An ultraintense hollow laser pulse---realizable, for instance, with a petawatt-class Laguerre--Gaussian (LG) beam---propagates in a near-critical-density plasma channel and drives a doughnut-shaped wakefield featuring an overcritical sheath front, which triggers longitudinal injection of downstream electrons into the wake bucket [Figs. \ref{fig1}(a) and (b)]. The injected high-energy electrons, counterpropagating in the laser fields, radiate a dense bath of MeV $\gamma$ photons via NCS while undergoing trajectory reversals within the wake, and are ultimately trapped at the rear wake by the longitudinal field. The trapped electron population, possessing substantial forward momenta, further emit and up-scatters forward-directed $\gamma$ photons through linear Compton scattering (LCS). It is within this anisotropic gamma-ray bath that frequent $\gamma\gamma$ collisions elevate the probability of LBW pair creation to a level sufficient for copious positron production inside the wake bucket [Figs. \ref{fig1}(c)--(e)]. Crucially, fraction of positrons generated within the frontier doughnut wakefield are immediately captured, focused, and accelerated, forming isolated attosecond-scale bunches with divergence angles $\theta_d \simeq 0.1~\mathrm{rad}$ and peak currents reaching hundreds of milliamperes [Fig. \ref{fig2}]. The combined action of NCS, LCS, and LBW processes mediates polarization transfer: circularly polarized $\gamma$ photons generated via LCS transfer their helicity to the nascent positrons through the LBW channel, endowing the bunches with significant net spin polarization [Fig. \ref{fig3}]. These phase-locked positrons gain energy to the GeV level in the longitudinal accelerating field [Fig. \ref{fig4}], and the robustness of the entire scheme is confirmed by parameter scans over laser intensity and plasma-channel density [Figs. \ref{fig5} and \ref{fig6}].

\section{Results of 3D and 2D PIC Simulations}

\subsection{Simulation setup and positron generation}
\begin{figure}[t!] 
	\begin{center}
		\includegraphics[width=\linewidth]{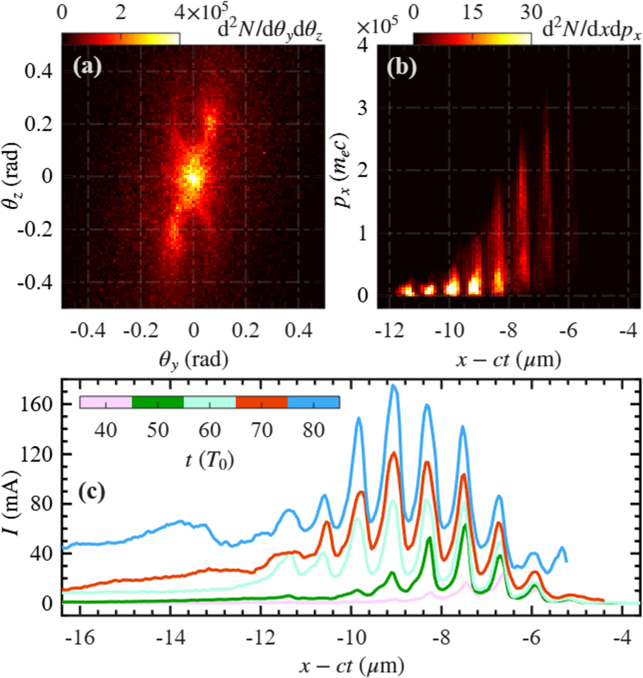}
		\caption{(a)-(b) Angular and phase-space distribution of accelerated positron bunch at time $t=58\:T_0$, where $\theta_y = {\rm arctan}(p_y/p_x)$ and $\theta_z = {\rm arctan}(p_z/p_x)$. (c) Current intensity of accelerated positron bunch at different times.}\label{fig2}
	\end{center}
\end{figure}

\begin{figure}[t!] 
	\begin{center}
		\includegraphics[width=\linewidth]{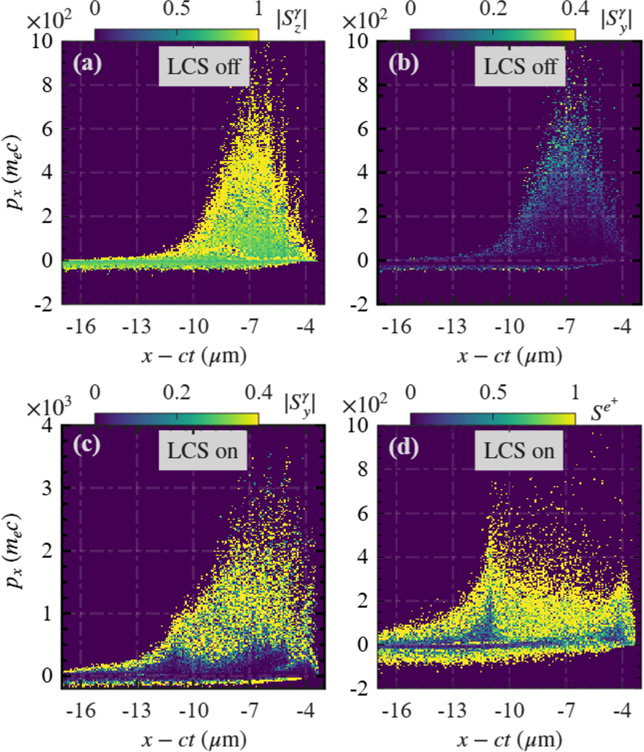}
		\caption{(a)-(b) Phase-space distributions of the linear polarization degree $|S_z^\gamma|$ and circular polarization degree $|S_y^\gamma|$ of gamma photons with the LCS deactivated (LCS off) at $t=58\:T_0$. (c)-(d) Phase-space distributions of the circular polarization degree $|S_y^\gamma|$ of gamma photons and the spin polarization $S$ of positrons newly produced via the LBW process with the LCS activated (LCS on) at $t=58\:T_0$.}\label{fig3}
	\end{center}
\end{figure}

The simulations employ SLIPs, a spin-resolved nonlinear--linear coupled QED particle-in-cell (PIC) code that models particle production and polarization transfer within a unified Monte Carlo framework \cite{Wan2023Simulations,dou2026Generation}. For nonlinear QED processes, spin-resolved event probabilities are evaluated under the locally constant field approximation using quasi-classical operator method \cite{sun2022Production}. For linear processes, the completely polarization and angle dependent crosssections are utilized \cite{zhao2023Cascade}. Between successive time steps, particle trajectories evolve according to the Lorentz equation, while spin precession is governed by the Thomas--Bargmann--Michel--Telegdi (TBMT) equation. The representative three-dimensional (3D) simulation configuration for capturing the full QED plasma physics and quantitatively studying positron bunches is as follows: a moving window propagates along the $+x$ direction, with a simulation domain of $15~\mu\mathrm{m} \times 20~\mu\mathrm{m} \times 20~\mu\mathrm{m}$ discretized into $300 \times 200 \times 200$ cells [Figs. \ref{fig1}(e), \ref{fig2}, \ref{fig3}, \ref{fig5}, and \ref{fig6}]. A more refined two-dimensional (2D) tracking simulation, with the z direction compressed, is used to qualitatively describe the particle dynamics; its simulation domain is 15 µm × 20 µm, discretized into 450 × 400 grid cells [Figs. \ref{fig1}(a)-\ref{fig1}(d) and \ref{fig4}]. A linearly polarized Laguerre--Gaussian laser pulse of wavelength $\lambda = 0.8~\mu\mathrm{m}$, spot size $w_0 = 3~\mu\mathrm{m}$, and peak normalized amplitude $a_0 \approx 266$ ($a_0 \approx 166$ for the 2D tracking simulation) drives a blowout wake in a preformed, fully ionized hydrocarbon plasma channel $[n_p(\mathrm{H}^+) = n_C(\mathrm{C}^{6+}) = n_e(e^-)/7]$. The transverse density profile follows $n_e = n_{\min} + (n_{\max} - n_{\min})(r/r_c)^5$, with $n_{\min} = 0.5\,n_c$, $n_{\max} = 2\,n_c$, $r = \sqrt{y^2 + z^2}$, and channel radius $r_c = 10~\mu\mathrm{m}$. Each cell is initialized with four macroelectrons, two macroprotons, and two macrocarbon ions. The normalizing parameters are the plasma critical density $n_c = \omega^2 m_e \epsilon_0 / e^2$, the laser period $T_0 = \lambda / c$, and the characteristic fields $E_0 = m_e c \omega / e$ and $B_0 = m_e \omega / e$, where $\omega = 2\pi c / \lambda$.

Such an ultraintense laser propagates through the near-critical-density plasma channel and drives the formation of a well-defined doughnut-shaped bubble structure [Fig. \ref{fig1}(a)]. The ponderomotive force of the laser simultaneously sweeps plasma electrons toward the bubble's leading edge, where they accumulate into a dense electron sheath. This sheath excites a longitudinal charge-separation field that, in turn, imprints a strong longitudinal modulation on the laser field [Fig. \ref{fig1}(b)]. A fraction of the downstream electrons at the bubble forefront are accelerated backward by this longitudinal field and injected into the bubble interior. Under the action of the laser's transverse electromagnetic field, the injected electrons execute oscillatory, wavelike trajectories within the cavity. At the turning points of these oscillations, the electron quantum parameter $\chi_{e^-}$ approaches or exceeds 0.1, dramatically enhancing the probability of NCS photon emission. As the electrons continue their motion through the wake, $\chi_{e^-}$ exhibits intermittent rises and falls, reflecting the competition between radiative energy loss and re-acceleration by the longitudinal electric field. During this process, a subset of electrons reverse their longitudinal direction and propagate forward while emitting forward-directed photons [Figs. \ref{fig1}(c) and 1(d)]. Notably, as electrons approach the rear half of the bubble where the longitudinal field is predominantly decelerating, the population of forward-moving electrons progressively increases. The net result is that electrons radiate $\gamma$-ray photons across a broad range of directions throughout their motion, collectively generating a dense, anisotropic gamma-ray bath of MeV-scale energy. Within this photon gas, $\gamma\gamma$ collisions occur at a rate sufficient to drive significant LBW pair production. Because the accelerating--focusing phases are periodically distributed along the longitudinal axis of the wakefield, positrons created at spatially separated locations are continuously captured into these favorable phases, assembling into a train of high-density positron bunches [Fig. \ref{fig1}(e)]. The sub-cycle positron bunches within the accelerating--focusing phases can be modulated with the peak density reaching about $10^{17}~\mathrm{cm}^{-3}$.

The coupled NCS--LCS--LBW interaction sequence established in Fig. \ref{fig1} culminates in the formation of a well-collimated attosecond positron bunch train, as characterized in Fig. \ref{fig2}. The angular distribution of the accelerated positrons reveals a discernible asymmetry in the divergence pattern [Fig. \ref{fig2}(a)], which reflects the intrinsic transverse asymmetry of LG laser field: the azimuthally non-uniform ponderomotive force imprints a correspondingly asymmetric transverse momentum distribution on the nascent positrons during their capture into the accelerating--focusing phases. In the longitudinal phase space, positrons phase-locked closer to the bubble leading edge acquire substantially higher Lorentz factors than those trapped farther rearward [Fig. \ref{fig2}(b)]. This energy gradient follows directly from the progressive strengthening of the longitudinal accelerating field $E_x$ toward the bubble front [Fig. \ref{fig1}(b)], which attains gradients of up to $80~\mathrm{TV/m}$ and preferentially energizes positrons captured in the forward portion of the wake. The temporal evolution of the bunch train further reveals that the positron population is predominantly concentrated in the trailing region of the pulse [Fig. \ref{fig2}(c)]. This spatial distribution arises because the LBW pair-production rate is maximized in the central volume of the plasma bubble, where the $\gamma$-photon density is highest [Fig. \ref{fig2}(c)]; positrons produced in this central region are subsequently overtaken by and captured into the decelerating--focusing phases at the rear of the wakefield, leading to a progressive accumulation of dark current toward the pulse tail [see Fig. \ref{fig4}(a) below]. As the laser propagates, the beam evolves from a few discrete bunches into a dense train of attosecond-scale pulses with peak currents exceeding several hundred milliamperes, each bunch trapped into a distinct accelerating--focusing phase.

\subsection{Polarization transfer and energy gain}
The spin polarization structure of the radiation bath and the resulting positron beam reveals a distinctive polarization transfer between nonlinear and linear QED processes. Gamma photons generated via NCS exhibit a high degree of linear polarization, quantified by $|S_L^\gamma| = \big|\sum_{i=1}^{N_{\gamma}} S_z^i\big|$, but negligible circular polarization $|S_C^\gamma| = \big|\sum_{i=1}^{N_{\gamma}} S_y^i\big|$ [Figs. \ref{fig3}(a) and \ref{fig3}(b)]. The LCS process, however, imprints a substantial degree of circular polarization on a fraction of the photon population [Figs. \ref{fig3}(b) and \ref{fig3}(c)]. This circular polarization is transferred to the nascent positrons through the LBW pair-production channel, endowing the positron beam with a net spin polarization $S^{e^+} = \sqrt{\big(\sum_{i=1}^{N_{e^+}} S_x^i\big)^2 + \big(\sum_{i=1}^{N_{e^+}} S_y^i\big)^2 + \big(\sum_{i=1}^{N_{e^+}} S_z^i\big)^2}$ [Fig. \ref{fig3}(d)]. This polarization transfer chain---NCS for photon production, LCS for circular-polarization imprinting, and LBW for helicity transfer to the pair---constitutes a self-consistent, all-optical polarization mechanism requiring no external spin-polarized drivers. 

\begin{figure}[t!] 
	\begin{center}
		\includegraphics[width=\linewidth]{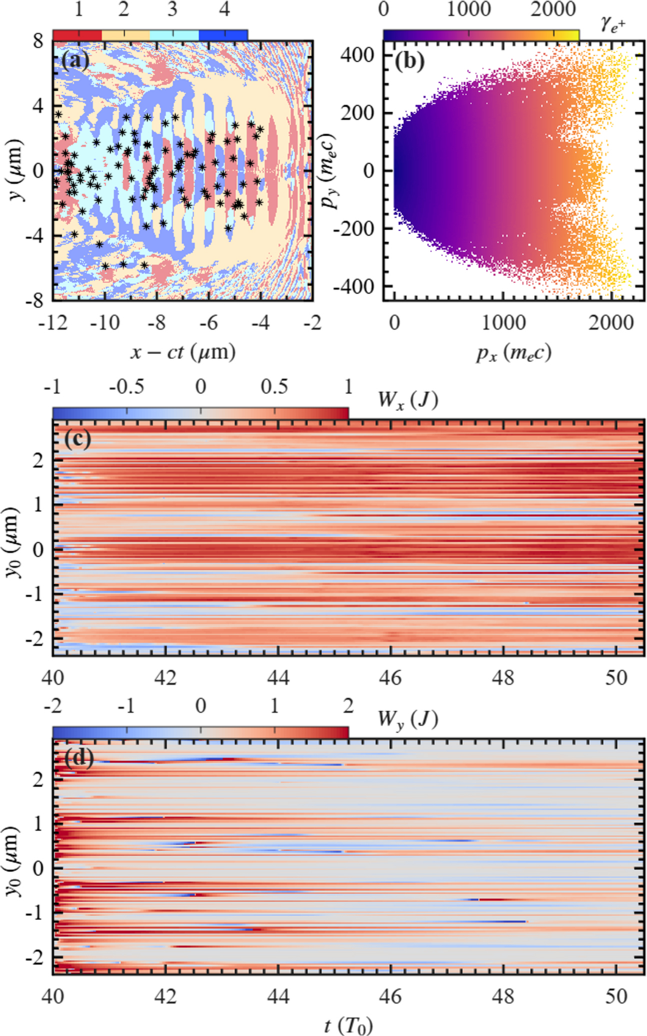}
		\caption{(a) Phase distribution of acceleration/deceleration and focusing/defocusing in the wakefield, 1 denotes the accelerating-focusing phase, 2 represents the accelerating-defocusing phase, 3 corresponds to the decelerating-focusing phase, and 4 indicates the decelerating-defocusing phase, at $t=40\:T_0$. The black asterisks represent the newly generated positrons randomly selected at this moment. (b) Phase-space distribution of the gamma factor $\gamma_{e^+}$ for accelerated and focused positrons at $t=50\:T_0$. (c)-(d) The longitudinal work $W_x$ and transverse work $W_y$ performed by the electromagnetic field on positrons initially located at different transverse positions $y_0$, as functions of time.}\label{fig4}
	\end{center}
\end{figure}
The bubble region for positrons can be divided into four distinct phases based on their longitudinal acceleration/deceleration and transverse focusing/defocusing behavior: (1) the acceleration-focusing phase ($E_x > 0$, $E_\perp < 0$ for $y > 0$, $E_\perp > 0$ for $y < 0$), (2) the acceleration-defocusing phase ($E_x > 0$, $E_\perp > 0$ for $y > 0$, $E_\perp < 0$ for $y < 0$), (3) the deceleration-focusing phase ($E_x < 0$, $E_\perp < 0$ for $y > 0$, $E_\perp > 0$ for $y < 0$), and (4) the deceleration-defocusing phase ($E_x < 0$, $E_\perp > 0$ for $y > 0$, $E_\perp < 0$ for $y < 0$) [Fig.~\ref{fig4}(a)], where $E_\perp = E_y - cB_z$. Positrons generated via photon collisions within the bubble populate these different phases and undergo distinct dynamical evolutions. Only those positrons located in the front half of the bubble have the potential to be trapped and accelerated to high energies. The accelerated positrons exhibit predominantly longitudinal momentum $p_x$ while maintaining non-negligible transverse momentum $p_y$ [Fig.~\ref{fig4}(b)]. The energy gain of trapped positrons can be illustrated by the evolutionary work of longitudinal and transverse fields for each trapped positron with a initial transverse position $y_0$ [Fig.~\ref{fig4}(c)].
This characteristic momenta distribution arises because the acceleration is primarily governed by longitudinal fields. However, when longitudinal acceleration weakens or transitions to deceleration, transverse acceleration becomes the dominant mechanism via magnetic-assisted direct laser acceleration [Fig.~\ref{fig4}(d)].

\begin{figure}[t] 
	\setlength{\abovecaptionskip}{-0.6cm}
	\begin{center}
		\includegraphics[width=\linewidth]{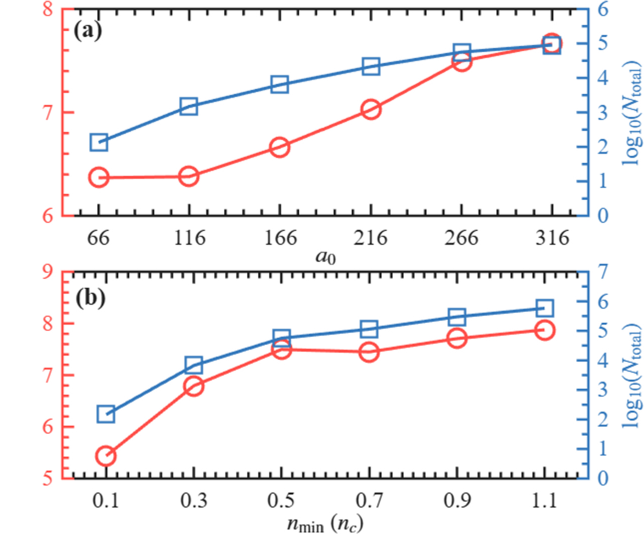}
		\begin{picture}(300,25)
			\put(1,160){\fontsize{9}{0}\selectfont\rotatebox{90}{\textcolor[RGB]{255,78,72}{$\rm{log_{10}}(\mathcal{B})\:(\rm{A/m^2})$}}}
			\put(1,60){\fontsize{9}{0}\selectfont\rotatebox{90}{\textcolor[RGB]{255,78,72}{$\rm{log_{10}}(\mathcal{B})\:(\rm{A/m^2})$}}}
		\end{picture}
		\caption{Effects of (a) laser intensity $a_0$ and (b) plasma channel density $n_{\rm min}$ on the brightness $\mathcal{B}$ and total number $N_{\rm total}$ of positrons. All data are taken at the time $t=58\:T_0$. Other parameters are the same as those in Fig. ~\ref{fig2}.}\label{fig5}
	\end{center}
\end{figure}
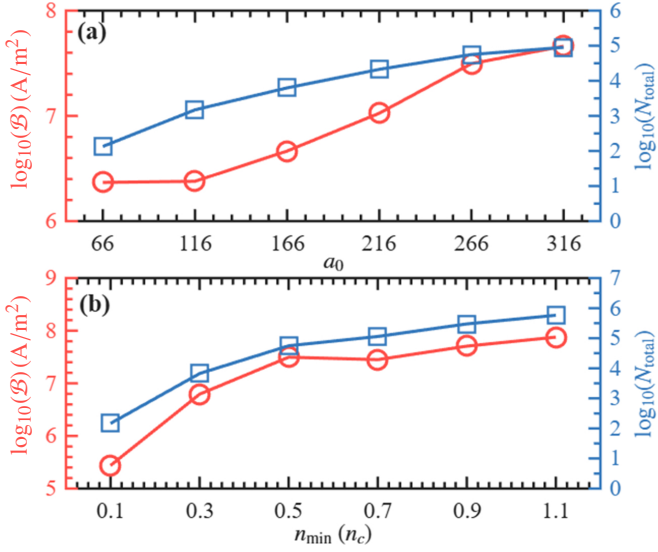
\subsection{Parameters Scan and Discussion}
\begin{figure}[t] 
	\begin{center}
		\includegraphics[width=\linewidth]{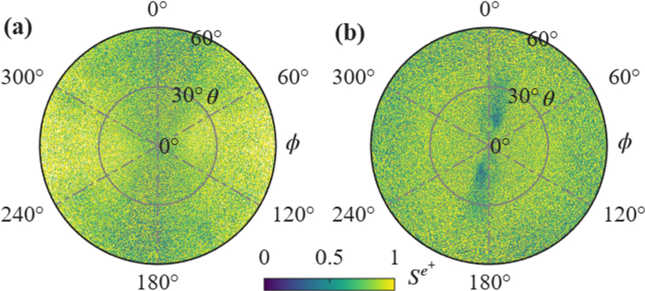}
		\caption{Distribution of the positron spin polarization $S^{e^+}$ in the $\theta-\phi$ plane of spherical coordinates at $t=58~T_0$: (a) $a_0=166$, (b) $a_0=266$. Other parameters are the same as those in Fig. ~\ref{fig2}.}\label{fig6}
	\end{center}
\end{figure}

The robustness of this \emph{in-situ} generation method is systematically verified by scanning the drive laser intensity $a_0$ and the minimum plasma-channel density $n_{\min}$. As $a_0$ increases, both the beam brightness $\mathcal{B}$ and the total positron yield $N_{\mathrm{total}}$ increase monotonically [Fig. \ref{fig5}(a)]. This trend follows directly from the scaling $\chi_{e^-} \propto E \propto a_0$: larger $a_0$ enhances the NCS photon emission rate from injected electrons, increasing the $\gamma$-photon density and thereby the LBW pair-production probability. Concurrently, stronger $a_0$ amplifies both the longitudinal accelerating field $E_x$ and the transverse focusing field $E_\perp$, improving the trapping efficiency and transverse confinement of the generated positrons. Meanwhile, increasing $n_{\min}$ also enhances $\mathcal{B}$, $N_{\mathrm{total}}$ [Fig. \ref{fig5}(b)]. A denser plasma channel accumulates more electrons at the bubble's leading edge, amplifying the charge-separation field and strengthening the longitudinal $E_x$, which in turn injects a larger population of higher-energy electrons into the bubble. These electrons radiate more copiously through both NCS and LCS channels, and the LCS-up-scattered photons, having both higher energy and greater circular polarization fraction, cross the LBW threshold more readily while transferring their helicity more efficiently to the produced pairs. Despite these enhancements in yield and brightness, however, the spin polarization of the positrons $S^{e^+}$ decreases significantly after a propagation distance of about $58~T_0$. This is because the precession frequency of the positron spins scales linearly with the local field amplitude; the stronger fields present during capture and acceleration induce more rapid spin precession, partially randomizing the spin orientations within the bunch  [Fig. \ref{fig6}].

Several theoretical schemes have been proposed for the experimental verification of the LBW process. These include the collision of two highly collimated $\gamma$-ray pulses driven by 10-PW lasers~\cite{yu2019Creation}; a single-laser approach in which a pulse traversing a dense plasma channel generates counterpropagating MeV photon populations via a longitudinal plasma electric field~\cite{he2021Single}; a laser--plasma platform combining LWFA-driven bremsstrahlung $\gamma$-rays with laser-heated plasma X-rays, developed at the Gemini facility~\cite{kettle2021Laser}; spin-resolved Monte Carlo investigations of the complete polarization signatures of the LBW process~\cite{zhao2022Signatures}; and a compact laser-driven scheme utilizing Thomson-scattered X-ray pulses~\cite{han2023Linear}. In all of these approaches, however, the generated positrons fail to form a high-density beam, rendering experimental detection exceedingly challenging owing to weak signal levels and the difficulty of discriminating the pair signal from overwhelming electromagnetic noise. In contrast, the radiation-wakefield-induced photon--photon collision mechanism demonstrated in the present work yields, for the first time, an LBW-generated positron beam with high density ($>10^{17}~\mathrm{cm}^{-3}$) and high current (several hundred milliamperes), thereby providing a distinctive and experimentally accessible signature for the unambiguous identification of the LBW process.

%Such polarized, collimated, high-current positron bunches open avenues for laboratory investigations of pulsar magnetospheres \cite{arrowsmith2024laboratory}, black-hole jets \cite{ruffini2010electron}, and gamma-ray burst physics \cite{kumar2015physics}, while the beam itself serves as a distinctive experimental signature of the LBW process, substantially reducing detection complexity.

\section{Summary}
We have proposed and numerically demonstrated a one-step, all-optical scheme for producing polarized attosecond positron beams via radiation-wakefield-induced two-photon pair creation. An ultraintense cylindrical-vector laser pulse propagating in a near-critical-density plasma channel drives a doughnut wakefield that injects electrons longitudinally; these electrons radiate a dense, quasi-isotropic bath of MeV $\gamma$ photons via nonlinear Compton scattering, and subsequent photon--photon collisions generate electron--positron pairs through the linear Breit--Wheeler process. Positrons born within the front half of the wakefield are immediately trapped, focused, and accelerated, completely bypassing the severe synchronization and transport losses inherent to external injection. Spin-resolved QED-PIC simulations reveal dense, collimated positron bunch trains with hundred-attosecond duration, GeV-scale energies, and significant net polarization originating from a self-consistent polarization-transfer chain linking nonlinear Compton scattering, linear Compton scattering, and linear Breit--Wheeler pair creation. The source performance is robust against practical variations in laser intensity and plasma-channel density, with denser channels simultaneously improving yield, brightness, and polarization. This compact, source-free mechanism establishes an accessible platform for strong-field QED studies, laboratory pair-plasma, and future polarized plasma-based accelerator concepts.

\vspace{1em}

{\it Acknowledgements}--- The work is supported by the National Natural Science Foundation of China (Grants No. 12425510, No. U2267204, No. 12441506, No. 12475249, No. 12447106, No. 12275209), the Science Challenge Project (No. TZ2025012), the National Key Research and Development (R\&D) Program (Grant No. 2024YFA1610900, No. 2024YFA1612700), the Innovative Scientific Program of CNNC, Natural Science Basic Research Program of Shaanxi (Grant No. 2024JC-YBQN-0042), and the Fundamental Research Funds for Central Universities (No. xzy012023046).

\bibliography{library}

\begin{thebibliography}{54}%
\makeatletter
\providecommand \@ifxundefined [1]{%
 \@ifx{#1\undefined}
}%
\providecommand \@ifnum [1]{%
 \ifnum #1\expandafter \@firstoftwo
 \else \expandafter \@secondoftwo
 \fi
}%
\providecommand \@ifx [1]{%
 \ifx #1\expandafter \@firstoftwo
 \else \expandafter \@secondoftwo
 \fi
}%
\providecommand \natexlab [1]{#1}%
\providecommand \enquote  [1]{``#1''}%
\providecommand \bibnamefont  [1]{#1}%
\providecommand \bibfnamefont [1]{#1}%
\providecommand \citenamefont [1]{#1}%
\providecommand \href@noop [0]{\@secondoftwo}%
\providecommand \href [0]{\begingroup \@sanitize@url \@href}%
\providecommand \@href[1]{\@@startlink{#1}\@@href}%
\providecommand \@@href[1]{\endgroup#1\@@endlink}%
\providecommand \@sanitize@url [0]{\catcode `\\12\catcode `\$12\catcode
  `\&12\catcode `\#12\catcode `\^12\catcode `\_12\catcode `\%12\relax}%
\providecommand \@@startlink[1]{}%
\providecommand \@@endlink[0]{}%
\providecommand \url  [0]{\begingroup\@sanitize@url \@url }%
\providecommand \@url [1]{\endgroup\@href {#1}{\urlprefix }}%
\providecommand \urlprefix  [0]{URL }%
\providecommand \Eprint [0]{\href }%
\providecommand \doibase [0]{https://doi.org/}%
\providecommand \selectlanguage [0]{\@gobble}%
\providecommand \bibinfo  [0]{\@secondoftwo}%
\providecommand \bibfield  [0]{\@secondoftwo}%
\providecommand \translation [1]{[#1]}%
\providecommand \BibitemOpen [0]{}%
\providecommand \bibitemStop [0]{}%
\providecommand \bibitemNoStop [0]{.\EOS\space}%
\providecommand \EOS [0]{\spacefactor3000\relax}%
\providecommand \BibitemShut  [1]{\csname bibitem#1\endcsname}%
\let\auto@bib@innerbib\@empty
%</preamble>
\bibitem [{\citenamefont {Danielson}\ \emph {et~al.}(2015)\citenamefont
  {Danielson}, \citenamefont {Dubin}, \citenamefont {Greaves},\ and\
  \citenamefont {Surko}}]{danielson2015Plasma}%
  \BibitemOpen
  \bibfield  {author} {\bibinfo {author} {\bibfnamefont {J.~R.}\ \bibnamefont
  {Danielson}}, \bibinfo {author} {\bibfnamefont {D.~H.~E.}\ \bibnamefont
  {Dubin}}, \bibinfo {author} {\bibfnamefont {R.~G.}\ \bibnamefont {Greaves}},\
  and\ \bibinfo {author} {\bibfnamefont {C.~M.}\ \bibnamefont {Surko}},\
  }\bibfield  {title} {\bibinfo {title} {Plasma and trap-based techniques for
  science with positrons},\ }\href {https://doi.org/10.1103/RevModPhys.87.247}
  {\bibfield  {journal} {\bibinfo  {journal} {Rev. Mod. Phys.}\ }\textbf
  {\bibinfo {volume} {87}},\ \bibinfo {pages} {247} (\bibinfo {year}
  {2015})}\BibitemShut {NoStop}%
\bibitem [{\citenamefont {An}\ \emph {et~al.}()\citenamefont {An},
  \citenamefont {Bai}, \citenamefont {Chen}, \citenamefont {Chen},
  \citenamefont {Chen},\ and\ \citenamefont {et~al}}]{an2019Precision}%
  \BibitemOpen
  \bibfield  {author} {\bibinfo {author} {\bibfnamefont {F.}~\bibnamefont
  {An}}, \bibinfo {author} {\bibfnamefont {Y.}~\bibnamefont {Bai}}, \bibinfo
  {author} {\bibfnamefont {C.}~\bibnamefont {Chen}}, \bibinfo {author}
  {\bibfnamefont {X.}~\bibnamefont {Chen}}, \bibinfo {author} {\bibfnamefont
  {Z.}~\bibnamefont {Chen}},\ and\ \bibinfo {author} {\bibnamefont {et~al}},\
  }\bibfield  {title} {\bibinfo {title} {Precision {{Higgs}} physics at the
  {{CEPC}} *},\ }\href {https://doi.org/10.1088/1674-1137/43/4/043002}
  {\bibfield  {journal} {\bibinfo  {journal} {Chin. Phys. C}\ }\textbf
  {\bibinfo {volume} {43}},\ \bibinfo {pages} {043002}}\BibitemShut {NoStop}%
\bibitem [{\citenamefont {Fajans}\ and\ \citenamefont
  {Surko}(2020)}]{fajans2020Plasma}%
  \BibitemOpen
  \bibfield  {author} {\bibinfo {author} {\bibfnamefont {J.}~\bibnamefont
  {Fajans}}\ and\ \bibinfo {author} {\bibfnamefont {C.~M.}\ \bibnamefont
  {Surko}},\ }\bibfield  {title} {\bibinfo {title} {Plasma and trap-based
  techniques for science with antimatter},\ }\href
  {https://doi.org/10.1063/1.5131273} {\bibfield  {journal} {\bibinfo
  {journal} {Phys. Plasmas}\ }\textbf {\bibinfo {volume} {27}},\ \bibinfo
  {pages} {030601} (\bibinfo {year} {2020})}\BibitemShut {NoStop}%
\bibitem [{\citenamefont {Chaikovska}\ \emph {et~al.}(2022)\citenamefont
  {Chaikovska}, \citenamefont {Chehab}, \citenamefont {Kubytskyi},
  \citenamefont {Ogur}, \citenamefont {Ushakov}, \citenamefont {Variola},
  \citenamefont {Sievers}, \citenamefont {Musumeci}, \citenamefont {Bandiera},
  \citenamefont {Enomoto}, \citenamefont {Hogan},\ and\ \citenamefont
  {Martyshkin}}]{chaikovska2022Positron}%
  \BibitemOpen
  \bibfield  {author} {\bibinfo {author} {\bibfnamefont {I.}~\bibnamefont
  {Chaikovska}}, \bibinfo {author} {\bibfnamefont {R.}~\bibnamefont {Chehab}},
  \bibinfo {author} {\bibfnamefont {V.}~\bibnamefont {Kubytskyi}}, \bibinfo
  {author} {\bibfnamefont {S.}~\bibnamefont {Ogur}}, \bibinfo {author}
  {\bibfnamefont {A.}~\bibnamefont {Ushakov}}, \bibinfo {author} {\bibfnamefont
  {A.}~\bibnamefont {Variola}}, \bibinfo {author} {\bibfnamefont
  {P.}~\bibnamefont {Sievers}}, \bibinfo {author} {\bibfnamefont
  {P.}~\bibnamefont {Musumeci}}, \bibinfo {author} {\bibfnamefont
  {L.}~\bibnamefont {Bandiera}}, \bibinfo {author} {\bibfnamefont
  {Y.}~\bibnamefont {Enomoto}}, \bibinfo {author} {\bibfnamefont
  {M.}~\bibnamefont {Hogan}},\ and\ \bibinfo {author} {\bibfnamefont
  {P.}~\bibnamefont {Martyshkin}},\ }\bibfield  {title} {\bibinfo {title}
  {Positron sources: From conventional to advanced accelerator concepts-based
  colliders},\ }\href {https://doi.org/10.1088/1748-0221/17/05/P05015}
  {\bibfield  {journal} {\bibinfo  {journal} {J. Instrum.}\ }\textbf {\bibinfo
  {volume} {17}}\bibinfo  {number} { (05)},\ \bibinfo {pages}
  {P05015}}\BibitemShut {NoStop}%
\bibitem [{\citenamefont {Cao}\ \emph {et~al.}(2024)\citenamefont {Cao},
  \citenamefont {Lindstr\o{}m}, \citenamefont {Adli}, \citenamefont {Corde},\
  and\ \citenamefont {Gessner}}]{cao2024Positron}%
  \BibitemOpen
\bibfield  {number} {  }\bibfield  {author} {\bibinfo {author} {\bibfnamefont
  {G.~J.}\ \bibnamefont {Cao}}, \bibinfo {author} {\bibfnamefont {C.~A.}\
  \bibnamefont {Lindstr\o{}m}}, \bibinfo {author} {\bibfnamefont
  {E.}~\bibnamefont {Adli}}, \bibinfo {author} {\bibfnamefont {S.}~\bibnamefont
  {Corde}},\ and\ \bibinfo {author} {\bibfnamefont {S.}~\bibnamefont
  {Gessner}},\ }\bibfield  {title} {\bibinfo {title} {Positron acceleration in
  plasma wakefields},\ }\href
  {https://doi.org/10.1103/PhysRevAccelBeams.27.034801} {\bibfield  {journal}
  {\bibinfo  {journal} {Phys. Rev. Accel. Beams}\ }\textbf {\bibinfo {volume}
  {27}},\ \bibinfo {pages} {034801} (\bibinfo {year} {2024})}\BibitemShut
  {NoStop}%
\bibitem [{\citenamefont {Si}\ and\ \citenamefont
  {Huang}(2024)}]{si2024Research}%
  \BibitemOpen
  \bibfield  {author} {\bibinfo {author} {\bibfnamefont {M.}~\bibnamefont
  {Si}}\ and\ \bibinfo {author} {\bibfnamefont {Y.}~\bibnamefont {Huang}},\
  }\bibfield  {title} {\bibinfo {title} {Research progress on advanced positron
  acceleration},\ }\href {https://doi.org/10.1140/epja/s10050-024-01433-0}
  {\bibfield  {journal} {\bibinfo  {journal} {Eur. Phys. J. A}\ }\textbf
  {\bibinfo {volume} {60}},\ \bibinfo {pages} {210} (\bibinfo {year}
  {2024})}\BibitemShut {NoStop}%
\bibitem [{\citenamefont {Ruffini}\ \emph {et~al.}(2010)\citenamefont
  {Ruffini}, \citenamefont {Vereshchagin},\ and\ \citenamefont
  {Xue}}]{ruffini2010electron}%
  \BibitemOpen
  \bibfield  {author} {\bibinfo {author} {\bibfnamefont {R.}~\bibnamefont
  {Ruffini}}, \bibinfo {author} {\bibfnamefont {G.}~\bibnamefont
  {Vereshchagin}},\ and\ \bibinfo {author} {\bibfnamefont {S.-S.}\ \bibnamefont
  {Xue}},\ }\bibfield  {title} {\bibinfo {title} {Electron--positron pairs in
  physics and astrophysics: {{From}} heavy nuclei to black holes},\ }\href
  {https://doi.org/10.1016/j.physrep.2009.10.004} {\bibfield  {journal}
  {\bibinfo  {journal} {Phys. Rep.}\ }\textbf {\bibinfo {volume} {487}},\
  \bibinfo {pages} {1} (\bibinfo {year} {2010})}\BibitemShut {NoStop}%
\bibitem [{\citenamefont {Kumar}\ and\ \citenamefont
  {Zhang}(2015)}]{kumar2015physics}%
  \BibitemOpen
  \bibfield  {author} {\bibinfo {author} {\bibfnamefont {P.}~\bibnamefont
  {Kumar}}\ and\ \bibinfo {author} {\bibfnamefont {B.}~\bibnamefont {Zhang}},\
  }\bibfield  {title} {\bibinfo {title} {The physics of gamma-ray bursts \&
  relativistic jets},\ }\href {https://doi.org/10.1016/j.physrep.2014.09.008}
  {\bibfield  {journal} {\bibinfo  {journal} {Phys. Rep.}\ }\bibinfo {series}
  {The Physics of Gamma-Ray Bursts \& Relativistic Jets},\ \textbf {\bibinfo
  {volume} {561}},\ \bibinfo {pages} {1} (\bibinfo {year} {2015})}\BibitemShut
  {NoStop}%
\bibitem [{\citenamefont {Sarri}\ \emph {et~al.}(2015)\citenamefont {Sarri},
  \citenamefont {Poder}, \citenamefont {Cole}, \citenamefont {Schumaker},
  \citenamefont {Di~Piazza},\ and\ \citenamefont
  {Reville}}]{sarri2015Generation}%
  \BibitemOpen
  \bibfield  {author} {\bibinfo {author} {\bibfnamefont {G.}~\bibnamefont
  {Sarri}}, \bibinfo {author} {\bibfnamefont {K.}~\bibnamefont {Poder}},
  \bibinfo {author} {\bibfnamefont {J.~M.}\ \bibnamefont {Cole}}, \bibinfo
  {author} {\bibfnamefont {W.}~\bibnamefont {Schumaker}}, \bibinfo {author}
  {\bibfnamefont {A.}~\bibnamefont {Di~Piazza}},\ and\ \bibinfo {author}
  {\bibfnamefont {B.~e.~a.}\ \bibnamefont {Reville}},\ }\bibfield  {title}
  {\bibinfo {title} {Generation of neutral and high-density electron–positron
  pair plasmas in the laboratory},\ }\href {https://doi.org/10.1038/ncomms7747}
  {\bibfield  {journal} {\bibinfo  {journal} {Nat. Commun.}\ }\textbf {\bibinfo
  {volume} {6}},\ \bibinfo {pages} {6747} (\bibinfo {year} {2015})}\BibitemShut
  {NoStop}%
\bibitem [{\citenamefont {Stoneking}\ \emph {et~al.}(2020)\citenamefont
  {Stoneking}, \citenamefont {Pedersen}, \citenamefont {Helander},\ and\
  \citenamefont {et~al}}]{stoneking2020new}%
  \BibitemOpen
  \bibfield  {author} {\bibinfo {author} {\bibfnamefont {M.~R.}\ \bibnamefont
  {Stoneking}}, \bibinfo {author} {\bibfnamefont {T.~S.}\ \bibnamefont
  {Pedersen}}, \bibinfo {author} {\bibfnamefont {P.}~\bibnamefont {Helander}},\
  and\ \bibinfo {author} {\bibnamefont {et~al}},\ }\bibfield  {title} {\bibinfo
  {title} {A new frontier in laboratory physics: Magnetized electron--positron
  plasmas},\ }\href {https://doi.org/10.1017/S0022377820001385} {\bibfield
  {journal} {\bibinfo  {journal} {J. Plasma Phys.}\ }\textbf {\bibinfo {volume}
  {86}},\ \bibinfo {pages} {155860601} (\bibinfo {year} {2020})}\BibitemShut
  {NoStop}%
\bibitem [{\citenamefont {Hada}\ \emph {et~al.}(2024)\citenamefont {Hada},
  \citenamefont {Asada}, \citenamefont {Nakamura},\ and\ \citenamefont
  {Kino}}]{hada2024M87}%
  \BibitemOpen
  \bibfield  {author} {\bibinfo {author} {\bibfnamefont {K.}~\bibnamefont
  {Hada}}, \bibinfo {author} {\bibfnamefont {K.}~\bibnamefont {Asada}},
  \bibinfo {author} {\bibfnamefont {M.}~\bibnamefont {Nakamura}},\ and\
  \bibinfo {author} {\bibfnamefont {M.}~\bibnamefont {Kino}},\ }\bibfield
  {title} {\bibinfo {title} {M 87: A cosmic laboratory for deciphering black
  hole accretion and jet formation},\ }\href
  {https://doi.org/10.1007/s00159-024-00155-y} {\bibfield  {journal} {\bibinfo
  {journal} {Astron. Astrophys. Rev.}\ }\textbf {\bibinfo {volume} {32}},\
  \bibinfo {pages} {5} (\bibinfo {year} {2024})}\BibitemShut {NoStop}%
\bibitem [{\citenamefont {Arrowsmith}\ \emph {et~al.}(2024)\citenamefont
  {Arrowsmith}, \citenamefont {Simon}, \citenamefont {Bilbao}, \citenamefont
  {Bott}, \citenamefont {Burger}, \citenamefont {Chen}, \citenamefont {Cruz},
  \citenamefont {Davenne}, \citenamefont {Efthymiopoulos}, \citenamefont
  {Froula}, \citenamefont {Goillot}, \citenamefont {Gudmundsson}, \citenamefont
  {Haberberger}, \citenamefont {Halliday}, \citenamefont {Hodge}, \citenamefont
  {Huffman}, \citenamefont {Iaquinta}, \citenamefont {Miniati}, \citenamefont
  {Reville}, \citenamefont {Sarkar}, \citenamefont {Schekochihin},
  \citenamefont {Silva}, \citenamefont {Simpson}, \citenamefont {Stergiou},
  \citenamefont {Trines}, \citenamefont {Vieu}, \citenamefont {Charitonidis},
  \citenamefont {Bingham},\ and\ \citenamefont
  {Gregori}}]{arrowsmith2024laboratory}%
  \BibitemOpen
  \bibfield  {author} {\bibinfo {author} {\bibfnamefont {C.~D.}\ \bibnamefont
  {Arrowsmith}}, \bibinfo {author} {\bibfnamefont {P.}~\bibnamefont {Simon}},
  \bibinfo {author} {\bibfnamefont {P.~J.}\ \bibnamefont {Bilbao}}, \bibinfo
  {author} {\bibfnamefont {A.~F.~A.}\ \bibnamefont {Bott}}, \bibinfo {author}
  {\bibfnamefont {S.}~\bibnamefont {Burger}}, \bibinfo {author} {\bibfnamefont
  {H.}~\bibnamefont {Chen}}, \bibinfo {author} {\bibfnamefont {F.~D.}\
  \bibnamefont {Cruz}}, \bibinfo {author} {\bibfnamefont {T.}~\bibnamefont
  {Davenne}}, \bibinfo {author} {\bibfnamefont {I.}~\bibnamefont
  {Efthymiopoulos}}, \bibinfo {author} {\bibfnamefont {D.~H.}\ \bibnamefont
  {Froula}}, \bibinfo {author} {\bibfnamefont {A.}~\bibnamefont {Goillot}},
  \bibinfo {author} {\bibfnamefont {J.~T.}\ \bibnamefont {Gudmundsson}},
  \bibinfo {author} {\bibfnamefont {D.}~\bibnamefont {Haberberger}}, \bibinfo
  {author} {\bibfnamefont {J.~W.~D.}\ \bibnamefont {Halliday}}, \bibinfo
  {author} {\bibfnamefont {T.}~\bibnamefont {Hodge}}, \bibinfo {author}
  {\bibfnamefont {B.~T.}\ \bibnamefont {Huffman}}, \bibinfo {author}
  {\bibfnamefont {S.}~\bibnamefont {Iaquinta}}, \bibinfo {author}
  {\bibfnamefont {F.}~\bibnamefont {Miniati}}, \bibinfo {author} {\bibfnamefont
  {B.}~\bibnamefont {Reville}}, \bibinfo {author} {\bibfnamefont
  {S.}~\bibnamefont {Sarkar}}, \bibinfo {author} {\bibfnamefont {A.~A.}\
  \bibnamefont {Schekochihin}}, \bibinfo {author} {\bibfnamefont {L.~O.}\
  \bibnamefont {Silva}}, \bibinfo {author} {\bibfnamefont {R.}~\bibnamefont
  {Simpson}}, \bibinfo {author} {\bibfnamefont {V.}~\bibnamefont {Stergiou}},
  \bibinfo {author} {\bibfnamefont {R.~M. G.~M.}\ \bibnamefont {Trines}},
  \bibinfo {author} {\bibfnamefont {T.}~\bibnamefont {Vieu}}, \bibinfo {author}
  {\bibfnamefont {N.}~\bibnamefont {Charitonidis}}, \bibinfo {author}
  {\bibfnamefont {R.}~\bibnamefont {Bingham}},\ and\ \bibinfo {author}
  {\bibfnamefont {G.}~\bibnamefont {Gregori}},\ }\bibfield  {title} {\bibinfo
  {title} {Laboratory realization of relativistic pair-plasma beams},\ }\href
  {https://doi.org/10.1038/s41467-024-49346-2} {\bibfield  {journal} {\bibinfo
  {journal} {Nat. Commun.}\ }\textbf {\bibinfo {volume} {15}},\ \bibinfo
  {pages} {5029} (\bibinfo {year} {2024})}\BibitemShut {NoStop}%
\bibitem [{\citenamefont {Abbott}\ \emph {et~al.}(2016)\citenamefont {Abbott},
  \citenamefont {Adderley}, \citenamefont {Adeyemi},\ and\ \citenamefont
  {et~al}}]{abbott2016Production}%
  \BibitemOpen
  \bibfield  {author} {\bibinfo {author} {\bibfnamefont {D.}~\bibnamefont
  {Abbott}}, \bibinfo {author} {\bibfnamefont {P.}~\bibnamefont {Adderley}},
  \bibinfo {author} {\bibfnamefont {A.}~\bibnamefont {Adeyemi}},\ and\ \bibinfo
  {author} {\bibnamefont {et~al}},\ }\bibfield  {title} {\bibinfo {title}
  {Production of {{Highly Polarized Positrons Using Polarized Electrons}} at
  {{MeV Energies}}},\ }\href {https://doi.org/10.1103/PhysRevLett.116.214801}
  {\bibfield  {journal} {\bibinfo  {journal} {Phys. Rev. Lett.}\ }\textbf
  {\bibinfo {volume} {116}},\ \bibinfo {pages} {214801} (\bibinfo {year}
  {2016})}\BibitemShut {NoStop}%
\bibitem [{\citenamefont {Boscolo}\ \emph {et~al.}(2020)\citenamefont
  {Boscolo}, \citenamefont {Antonelli}, \citenamefont {Ciarma},\ and\
  \citenamefont {Raimondi}}]{boscolo2020Muon}%
  \BibitemOpen
  \bibfield  {author} {\bibinfo {author} {\bibfnamefont {M.}~\bibnamefont
  {Boscolo}}, \bibinfo {author} {\bibfnamefont {M.}~\bibnamefont {Antonelli}},
  \bibinfo {author} {\bibfnamefont {A.}~\bibnamefont {Ciarma}},\ and\ \bibinfo
  {author} {\bibfnamefont {P.}~\bibnamefont {Raimondi}},\ }\bibfield  {title}
  {\bibinfo {title} {Muon production and accumulation from positrons on
  target},\ }\href {https://doi.org/10.1103/PhysRevAccelBeams.23.051001}
  {\bibfield  {journal} {\bibinfo  {journal} {Phys. Rev. Accel. Beams}\
  }\textbf {\bibinfo {volume} {23}},\ \bibinfo {pages} {051001} (\bibinfo
  {year} {2020})}\BibitemShut {NoStop}%
\bibitem [{\citenamefont {Dou}\ \emph {et~al.}(2025)\citenamefont {Dou},
  \citenamefont {Lv}, \citenamefont {Salamin}, \citenamefont {Zhang},
  \citenamefont {Wan}, \citenamefont {Xu},\ and\ \citenamefont
  {Li}}]{dou2025Compact}%
  \BibitemOpen
  \bibfield  {author} {\bibinfo {author} {\bibfnamefont {Z.-K.}\ \bibnamefont
  {Dou}}, \bibinfo {author} {\bibfnamefont {C.}~\bibnamefont {Lv}}, \bibinfo
  {author} {\bibfnamefont {Y.~I.}\ \bibnamefont {Salamin}}, \bibinfo {author}
  {\bibfnamefont {N.}~\bibnamefont {Zhang}}, \bibinfo {author} {\bibfnamefont
  {F.}~\bibnamefont {Wan}}, \bibinfo {author} {\bibfnamefont {Z.-F.}\
  \bibnamefont {Xu}},\ and\ \bibinfo {author} {\bibfnamefont {J.-X.}\
  \bibnamefont {Li}},\ }\bibfield  {title} {\bibinfo {title} {Compact
  spin-polarized positron acceleration in multilayer microhole-array films},\
  }\href {https://doi.org/10.1103/PhysRevE.111.035209} {\bibfield  {journal}
  {\bibinfo  {journal} {Phys. Rev. E}\ }\textbf {\bibinfo {volume} {111}},\
  \bibinfo {pages} {035209} (\bibinfo {year} {2025})}\BibitemShut {NoStop}%
\bibitem [{\citenamefont {Alejo}\ \emph {et~al.}(2019)\citenamefont {Alejo},
  \citenamefont {Walczak},\ and\ \citenamefont {Sarri}}]{alejo2019Laser}%
  \BibitemOpen
  \bibfield  {author} {\bibinfo {author} {\bibfnamefont {A.}~\bibnamefont
  {Alejo}}, \bibinfo {author} {\bibfnamefont {R.}~\bibnamefont {Walczak}},\
  and\ \bibinfo {author} {\bibfnamefont {G.}~\bibnamefont {Sarri}},\ }\bibfield
   {title} {\bibinfo {title} {Laser-driven high-quality positron sources as
  possible injectors for plasma-based accelerators},\ }\href
  {https://doi.org/10.1038/s41598-019-41650-y} {\bibfield  {journal} {\bibinfo
  {journal} {Sci. Rep.}\ }\textbf {\bibinfo {volume} {9}},\ \bibinfo {pages}
  {5279} (\bibinfo {year} {2019})}\BibitemShut {NoStop}%
\bibitem [{\citenamefont {Streeter}\ \emph {et~al.}(2024)\citenamefont
  {Streeter}, \citenamefont {Colgan}, \citenamefont {Carderelli}, \citenamefont
  {Ma}, \citenamefont {Cavanagh}, \citenamefont {Los}, \citenamefont {Ahmed},
  \citenamefont {Antoine}, \citenamefont {Audet}, \citenamefont {Balcazar},
  \citenamefont {Calvin}, \citenamefont {Kettle}, \citenamefont {Mangles},
  \citenamefont {Najmudin}, \citenamefont {Rajeev}, \citenamefont {Symes},
  \citenamefont {Thomas},\ and\ \citenamefont {Sarri}}]{streeter2024Narrow}%
  \BibitemOpen
  \bibfield  {author} {\bibinfo {author} {\bibfnamefont {M.~J.~V.}\
  \bibnamefont {Streeter}}, \bibinfo {author} {\bibfnamefont {C.}~\bibnamefont
  {Colgan}}, \bibinfo {author} {\bibfnamefont {J.}~\bibnamefont {Carderelli}},
  \bibinfo {author} {\bibfnamefont {Y.}~\bibnamefont {Ma}}, \bibinfo {author}
  {\bibfnamefont {N.}~\bibnamefont {Cavanagh}}, \bibinfo {author}
  {\bibfnamefont {E.~E.}\ \bibnamefont {Los}}, \bibinfo {author} {\bibfnamefont
  {H.}~\bibnamefont {Ahmed}}, \bibinfo {author} {\bibfnamefont {A.~F.}\
  \bibnamefont {Antoine}}, \bibinfo {author} {\bibfnamefont {T.}~\bibnamefont
  {Audet}}, \bibinfo {author} {\bibfnamefont {M.~D.}\ \bibnamefont {Balcazar}},
  \bibinfo {author} {\bibfnamefont {L.}~\bibnamefont {Calvin}}, \bibinfo
  {author} {\bibfnamefont {B.}~\bibnamefont {Kettle}}, \bibinfo {author}
  {\bibfnamefont {S.~P.~D.}\ \bibnamefont {Mangles}}, \bibinfo {author}
  {\bibfnamefont {Z.}~\bibnamefont {Najmudin}}, \bibinfo {author}
  {\bibfnamefont {P.~P.}\ \bibnamefont {Rajeev}}, \bibinfo {author}
  {\bibfnamefont {D.~R.}\ \bibnamefont {Symes}}, \bibinfo {author}
  {\bibfnamefont {A.~G.~R.}\ \bibnamefont {Thomas}},\ and\ \bibinfo {author}
  {\bibfnamefont {G.}~\bibnamefont {Sarri}},\ }\bibfield  {title} {\bibinfo
  {title} {Narrow bandwidth, low-emittance positron beams from a
  laser-wakefield accelerator},\ }\href
  {https://doi.org/10.1038/s41598-024-56281-1} {\bibfield  {journal} {\bibinfo
  {journal} {Sci. Rep.}\ }\textbf {\bibinfo {volume} {14}},\ \bibinfo {pages}
  {6001} (\bibinfo {year} {2024})}\BibitemShut {NoStop}%
\bibitem [{\citenamefont {Hessami}\ and\ \citenamefont
  {Gessner}(2023)}]{hessami2023Compact}%
  \BibitemOpen
  \bibfield  {author} {\bibinfo {author} {\bibfnamefont {R.}~\bibnamefont
  {Hessami}}\ and\ \bibinfo {author} {\bibfnamefont {S.}~\bibnamefont
  {Gessner}},\ }\bibfield  {title} {\bibinfo {title} {Compact source of
  positron beams with small thermal emittance},\ }\href
  {https://doi.org/10.1103/PhysRevAccelBeams.26.123402} {\bibfield  {journal}
  {\bibinfo  {journal} {Phys. Rev. Accel. Beams}\ }\textbf {\bibinfo {volume}
  {26}},\ \bibinfo {pages} {123402} (\bibinfo {year} {2023})}\BibitemShut
  {NoStop}%
\bibitem [{\citenamefont {Sarri}\ \emph {et~al.}(2013)\citenamefont {Sarri},
  \citenamefont {Schumaker}, \citenamefont {Di~Piazza}, \citenamefont {Vargas},
  \citenamefont {Dromey}, \citenamefont {Dieckmann}, \citenamefont {Chvykov},
  \citenamefont {Maksimchuk}, \citenamefont {Yanovsky}, \citenamefont {He},
  \citenamefont {Hou}, \citenamefont {Nees}, \citenamefont {Thomas},
  \citenamefont {Keitel}, \citenamefont {Zepf},\ and\ \citenamefont
  {Krushelnick}}]{sarri2013TableTop}%
  \BibitemOpen
  \bibfield  {author} {\bibinfo {author} {\bibfnamefont {G.}~\bibnamefont
  {Sarri}}, \bibinfo {author} {\bibfnamefont {W.}~\bibnamefont {Schumaker}},
  \bibinfo {author} {\bibfnamefont {A.}~\bibnamefont {Di~Piazza}}, \bibinfo
  {author} {\bibfnamefont {M.}~\bibnamefont {Vargas}}, \bibinfo {author}
  {\bibfnamefont {B.}~\bibnamefont {Dromey}}, \bibinfo {author} {\bibfnamefont
  {M.~E.}\ \bibnamefont {Dieckmann}}, \bibinfo {author} {\bibfnamefont
  {V.}~\bibnamefont {Chvykov}}, \bibinfo {author} {\bibfnamefont
  {A.}~\bibnamefont {Maksimchuk}}, \bibinfo {author} {\bibfnamefont
  {V.}~\bibnamefont {Yanovsky}}, \bibinfo {author} {\bibfnamefont {Z.~H.}\
  \bibnamefont {He}}, \bibinfo {author} {\bibfnamefont {B.~X.}\ \bibnamefont
  {Hou}}, \bibinfo {author} {\bibfnamefont {J.~A.}\ \bibnamefont {Nees}},
  \bibinfo {author} {\bibfnamefont {A.~G.~R.}\ \bibnamefont {Thomas}}, \bibinfo
  {author} {\bibfnamefont {C.~H.}\ \bibnamefont {Keitel}}, \bibinfo {author}
  {\bibfnamefont {M.}~\bibnamefont {Zepf}},\ and\ \bibinfo {author}
  {\bibfnamefont {K.}~\bibnamefont {Krushelnick}},\ }\bibfield  {title}
  {\bibinfo {title} {Table-{{Top Laser-Based Source}} of {{Femtosecond}},
  {{Collimated}}, {{Ultrarelativistic Positron Beams}}},\ }\href
  {https://doi.org/10.1103/PhysRevLett.110.255002} {\bibfield  {journal}
  {\bibinfo  {journal} {Phys. Rev. Lett.}\ }\textbf {\bibinfo {volume} {110}},\
  \bibinfo {pages} {255002} (\bibinfo {year} {2013})}\BibitemShut {NoStop}%
\bibitem [{\citenamefont {Chen}\ \emph {et~al.}(2015)\citenamefont {Chen},
  \citenamefont {Fiuza}, \citenamefont {Link}, \citenamefont {Hazi},
  \citenamefont {Hill}, \citenamefont {Hoarty}, \citenamefont {James},
  \citenamefont {Kerr}, \citenamefont {Meyerhofer}, \citenamefont {Myatt},
  \citenamefont {Park}, \citenamefont {Sentoku},\ and\ \citenamefont
  {Williams}}]{chen2015Scaling}%
  \BibitemOpen
  \bibfield  {author} {\bibinfo {author} {\bibfnamefont {H.}~\bibnamefont
  {Chen}}, \bibinfo {author} {\bibfnamefont {F.}~\bibnamefont {Fiuza}},
  \bibinfo {author} {\bibfnamefont {A.}~\bibnamefont {Link}}, \bibinfo {author}
  {\bibfnamefont {A.}~\bibnamefont {Hazi}}, \bibinfo {author} {\bibfnamefont
  {M.}~\bibnamefont {Hill}}, \bibinfo {author} {\bibfnamefont {D.}~\bibnamefont
  {Hoarty}}, \bibinfo {author} {\bibfnamefont {S.}~\bibnamefont {James}},
  \bibinfo {author} {\bibfnamefont {S.}~\bibnamefont {Kerr}}, \bibinfo {author}
  {\bibfnamefont {D.~D.}\ \bibnamefont {Meyerhofer}}, \bibinfo {author}
  {\bibfnamefont {J.}~\bibnamefont {Myatt}}, \bibinfo {author} {\bibfnamefont
  {J.}~\bibnamefont {Park}}, \bibinfo {author} {\bibfnamefont {Y.}~\bibnamefont
  {Sentoku}},\ and\ \bibinfo {author} {\bibfnamefont {G.~J.}\ \bibnamefont
  {Williams}},\ }\bibfield  {title} {\bibinfo {title} {Scaling the yield of
  laser-driven electron-positron jets to laboratory astrophysical
  applications},\ }\href {https://doi.org/10.1103/PhysRevLett.114.215001}
  {\bibfield  {journal} {\bibinfo  {journal} {Phys. Rev. Lett.}\ }\textbf
  {\bibinfo {volume} {114}},\ \bibinfo {pages} {215001} (\bibinfo {year}
  {2015})}\BibitemShut {NoStop}%
\bibitem [{\citenamefont {Wu}\ \emph {et~al.}(2016)\citenamefont {Wu},
  \citenamefont {Han}, \citenamefont {Zhang}, \citenamefont {Dong},
  \citenamefont {Zhu}, \citenamefont {Yan},\ and\ \citenamefont
  {Gu}}]{wu2016Optimization}%
  \BibitemOpen
  \bibfield  {author} {\bibinfo {author} {\bibfnamefont {Y.}~\bibnamefont
  {Wu}}, \bibinfo {author} {\bibfnamefont {D.}~\bibnamefont {Han}}, \bibinfo
  {author} {\bibfnamefont {T.}~\bibnamefont {Zhang}}, \bibinfo {author}
  {\bibfnamefont {K.}~\bibnamefont {Dong}}, \bibinfo {author} {\bibfnamefont
  {B.}~\bibnamefont {Zhu}}, \bibinfo {author} {\bibfnamefont {Y.}~\bibnamefont
  {Yan}},\ and\ \bibinfo {author} {\bibfnamefont {Y.}~\bibnamefont {Gu}},\
  }\bibfield  {title} {\bibinfo {title} {Optimization of positrons generation
  based on laser wakefield electron acceleration},\ }\href
  {https://doi.org/10.1103/PhysRevAccelBeams.19.081303} {\bibfield  {journal}
  {\bibinfo  {journal} {Phys. Rev. Accel. Beams}\ }\textbf {\bibinfo {volume}
  {19}},\ \bibinfo {pages} {081303} (\bibinfo {year} {2016})}\BibitemShut
  {NoStop}%
\bibitem [{\citenamefont {Chen}\ and\ \citenamefont
  {Fiuza}(2023)}]{chen2023Perspectives}%
  \BibitemOpen
  \bibfield  {author} {\bibinfo {author} {\bibfnamefont {H.}~\bibnamefont
  {Chen}}\ and\ \bibinfo {author} {\bibfnamefont {F.}~\bibnamefont {Fiuza}},\
  }\bibfield  {title} {\bibinfo {title} {Perspectives on relativistic
  electron–positron pair plasma experiments of astrophysical relevance using
  high-power lasers},\ }\href {https://doi.org/10.1063/5.0134819} {\bibfield
  {journal} {\bibinfo  {journal} {Phys. Plasmas}\ }\textbf {\bibinfo {volume}
  {30}},\ \bibinfo {pages} {020601} (\bibinfo {year} {2023})}\BibitemShut
  {NoStop}%
\bibitem [{\citenamefont {Danson}\ \emph {et~al.}(2019)\citenamefont {Danson},
  \citenamefont {Haefner}, \citenamefont {Bromage}, \citenamefont {Butcher},
  \citenamefont {Chanteloup}, \citenamefont {Chowdhury}, \citenamefont
  {Galvanauskas}, \citenamefont {Gizzi}, \citenamefont {Hein}, \citenamefont
  {Hillier}, \citenamefont {Hopps}, \citenamefont {Kato}, \citenamefont
  {Khazanov}, \citenamefont {Kodama}, \citenamefont {Korn}, \citenamefont {Li},
  \citenamefont {Li}, \citenamefont {Limpert}, \citenamefont {Ma},
  \citenamefont {Nam}, \citenamefont {Neely}, \citenamefont {Papadopoulos},
  \citenamefont {Penman}, \citenamefont {Qian}, \citenamefont {Rocca},
  \citenamefont {Shaykin}, \citenamefont {Siders}, \citenamefont {Spindloe},
  \citenamefont {Szatm{\'a}ri}, \citenamefont {Trines}, \citenamefont {Zhu},
  \citenamefont {Zhu},\ and\ \citenamefont {Zuegel}}]{danson2019Petawatt}%
  \BibitemOpen
  \bibfield  {author} {\bibinfo {author} {\bibfnamefont {C.~N.}\ \bibnamefont
  {Danson}}, \bibinfo {author} {\bibfnamefont {C.}~\bibnamefont {Haefner}},
  \bibinfo {author} {\bibfnamefont {J.}~\bibnamefont {Bromage}}, \bibinfo
  {author} {\bibfnamefont {T.}~\bibnamefont {Butcher}}, \bibinfo {author}
  {\bibfnamefont {J.-C.~F.}\ \bibnamefont {Chanteloup}}, \bibinfo {author}
  {\bibfnamefont {E.~A.}\ \bibnamefont {Chowdhury}}, \bibinfo {author}
  {\bibfnamefont {A.}~\bibnamefont {Galvanauskas}}, \bibinfo {author}
  {\bibfnamefont {L.~A.}\ \bibnamefont {Gizzi}}, \bibinfo {author}
  {\bibfnamefont {J.}~\bibnamefont {Hein}}, \bibinfo {author} {\bibfnamefont
  {D.~I.}\ \bibnamefont {Hillier}}, \bibinfo {author} {\bibfnamefont {N.~W.}\
  \bibnamefont {Hopps}}, \bibinfo {author} {\bibfnamefont {Y.}~\bibnamefont
  {Kato}}, \bibinfo {author} {\bibfnamefont {E.~A.}\ \bibnamefont {Khazanov}},
  \bibinfo {author} {\bibfnamefont {R.}~\bibnamefont {Kodama}}, \bibinfo
  {author} {\bibfnamefont {G.}~\bibnamefont {Korn}}, \bibinfo {author}
  {\bibfnamefont {R.}~\bibnamefont {Li}}, \bibinfo {author} {\bibfnamefont
  {Y.}~\bibnamefont {Li}}, \bibinfo {author} {\bibfnamefont {J.}~\bibnamefont
  {Limpert}}, \bibinfo {author} {\bibfnamefont {J.}~\bibnamefont {Ma}},
  \bibinfo {author} {\bibfnamefont {C.~H.}\ \bibnamefont {Nam}}, \bibinfo
  {author} {\bibfnamefont {D.}~\bibnamefont {Neely}}, \bibinfo {author}
  {\bibfnamefont {D.}~\bibnamefont {Papadopoulos}}, \bibinfo {author}
  {\bibfnamefont {R.~R.}\ \bibnamefont {Penman}}, \bibinfo {author}
  {\bibfnamefont {L.}~\bibnamefont {Qian}}, \bibinfo {author} {\bibfnamefont
  {J.~J.}\ \bibnamefont {Rocca}}, \bibinfo {author} {\bibfnamefont {A.~A.}\
  \bibnamefont {Shaykin}}, \bibinfo {author} {\bibfnamefont {C.~W.}\
  \bibnamefont {Siders}}, \bibinfo {author} {\bibfnamefont {C.}~\bibnamefont
  {Spindloe}}, \bibinfo {author} {\bibfnamefont {S.}~\bibnamefont
  {Szatm{\'a}ri}}, \bibinfo {author} {\bibfnamefont {R.~M. G.~M.}\ \bibnamefont
  {Trines}}, \bibinfo {author} {\bibfnamefont {J.}~\bibnamefont {Zhu}},
  \bibinfo {author} {\bibfnamefont {P.}~\bibnamefont {Zhu}},\ and\ \bibinfo
  {author} {\bibfnamefont {J.~D.}\ \bibnamefont {Zuegel}},\ }\bibfield  {title}
  {\bibinfo {title} {Petawatt and exawatt class lasers worldwide},\ }\href
  {https://doi.org/10.1017/hpl.2019.36} {\bibfield  {journal} {\bibinfo
  {journal} {High Power Laser Sci. Eng.}\ }\textbf {\bibinfo {volume} {7}},\
  \bibinfo {pages} {e54} (\bibinfo {year} {2019})}\BibitemShut {NoStop}%
\bibitem [{\citenamefont {Yoon}\ \emph {et~al.}(2021)\citenamefont {Yoon},
  \citenamefont {Kim}, \citenamefont {Choi}, \citenamefont {Sung},
  \citenamefont {Lee}, \citenamefont {Lee},\ and\ \citenamefont
  {Nam}}]{yoon2021Realization}%
  \BibitemOpen
  \bibfield  {author} {\bibinfo {author} {\bibfnamefont {J.~W.}\ \bibnamefont
  {Yoon}}, \bibinfo {author} {\bibfnamefont {Y.~G.}\ \bibnamefont {Kim}},
  \bibinfo {author} {\bibfnamefont {I.~W.}\ \bibnamefont {Choi}}, \bibinfo
  {author} {\bibfnamefont {J.~H.}\ \bibnamefont {Sung}}, \bibinfo {author}
  {\bibfnamefont {H.~W.}\ \bibnamefont {Lee}}, \bibinfo {author} {\bibfnamefont
  {S.~K.}\ \bibnamefont {Lee}},\ and\ \bibinfo {author} {\bibfnamefont {C.~H.}\
  \bibnamefont {Nam}},\ }\bibfield  {title} {\bibinfo {title} {Realization of
  laser intensity over $10^{23}$ {{W}}/cm$^2$},\ }\href
  {https://doi.org/10.1364/OPTICA.420520} {\bibfield  {journal} {\bibinfo
  {journal} {Optica}\ }\textbf {\bibinfo {volume} {8}},\ \bibinfo {pages} {630}
  (\bibinfo {year} {2021})}\BibitemShut {NoStop}%
\bibitem [{\citenamefont {Yamanouchi}\ \emph {et~al.}(2021)\citenamefont
  {Yamanouchi}, \citenamefont {Midorikawa},\ and\ \citenamefont
  {Roso}}]{yamanouchi2021Progress}%
  \BibitemOpen
  \bibinfo {editor} {\bibfnamefont {K.}~\bibnamefont {Yamanouchi}}, \bibinfo
  {editor} {\bibfnamefont {K.}~\bibnamefont {Midorikawa}},\ and\ \bibinfo
  {editor} {\bibfnamefont {L.}~\bibnamefont {Roso}},\ eds.,\ \href
  {https://doi.org/10.1007/978-3-030-75089-3} {\emph {\bibinfo {title}
  {Progress in {{Ultrafast Intense Laser Science XVI}}}}},\ \bibinfo {series}
  {Topics in {{Applied Physics}}}, Vol.\ \bibinfo {volume} {141}\ (\bibinfo
  {publisher} {Springer International Publishing},\ \bibinfo {address} {Cham},\
  \bibinfo {year} {2021})\BibitemShut {NoStop}%
\bibitem [{\citenamefont {Radier}\ \emph {et~al.}(2022)\citenamefont {Radier},
  \citenamefont {Chalus}, \citenamefont {Charbonneau}, \citenamefont
  {Thambirajah}, \citenamefont {Deschamps}, \citenamefont {David},
  \citenamefont {Barbe}, \citenamefont {Etter}, \citenamefont {Matras},
  \citenamefont {Ricaud}, \citenamefont {Leroux}, \citenamefont {Richard},
  \citenamefont {Lureau}, \citenamefont {Baleanu}, \citenamefont {Banici},
  \citenamefont {Gradinariu}, \citenamefont {Caldararu}, \citenamefont
  {Capiteanu}, \citenamefont {Naziru}, \citenamefont {Diaconescu},
  \citenamefont {Iancu}, \citenamefont {Dabu}, \citenamefont {Ursescu},
  \citenamefont {Dancus}, \citenamefont {Ur}, \citenamefont {Tanaka},\ and\
  \citenamefont {Zamfir}}]{radier2022Peak}%
  \BibitemOpen
  \bibfield  {author} {\bibinfo {author} {\bibfnamefont {C.}~\bibnamefont
  {Radier}}, \bibinfo {author} {\bibfnamefont {O.}~\bibnamefont {Chalus}},
  \bibinfo {author} {\bibfnamefont {M.}~\bibnamefont {Charbonneau}}, \bibinfo
  {author} {\bibfnamefont {S.}~\bibnamefont {Thambirajah}}, \bibinfo {author}
  {\bibfnamefont {G.}~\bibnamefont {Deschamps}}, \bibinfo {author}
  {\bibfnamefont {S.}~\bibnamefont {David}}, \bibinfo {author} {\bibfnamefont
  {J.}~\bibnamefont {Barbe}}, \bibinfo {author} {\bibfnamefont
  {E.}~\bibnamefont {Etter}}, \bibinfo {author} {\bibfnamefont
  {G.}~\bibnamefont {Matras}}, \bibinfo {author} {\bibfnamefont
  {S.}~\bibnamefont {Ricaud}}, \bibinfo {author} {\bibfnamefont
  {V.}~\bibnamefont {Leroux}}, \bibinfo {author} {\bibfnamefont
  {C.}~\bibnamefont {Richard}}, \bibinfo {author} {\bibfnamefont
  {F.}~\bibnamefont {Lureau}}, \bibinfo {author} {\bibfnamefont
  {A.}~\bibnamefont {Baleanu}}, \bibinfo {author} {\bibfnamefont
  {R.}~\bibnamefont {Banici}}, \bibinfo {author} {\bibfnamefont
  {A.}~\bibnamefont {Gradinariu}}, \bibinfo {author} {\bibfnamefont
  {C.}~\bibnamefont {Caldararu}}, \bibinfo {author} {\bibfnamefont
  {C.}~\bibnamefont {Capiteanu}}, \bibinfo {author} {\bibfnamefont
  {A.}~\bibnamefont {Naziru}}, \bibinfo {author} {\bibfnamefont
  {B.}~\bibnamefont {Diaconescu}}, \bibinfo {author} {\bibfnamefont
  {V.}~\bibnamefont {Iancu}}, \bibinfo {author} {\bibfnamefont
  {R.}~\bibnamefont {Dabu}}, \bibinfo {author} {\bibfnamefont {D.}~\bibnamefont
  {Ursescu}}, \bibinfo {author} {\bibfnamefont {I.}~\bibnamefont {Dancus}},
  \bibinfo {author} {\bibfnamefont {C.~A.}\ \bibnamefont {Ur}}, \bibinfo
  {author} {\bibfnamefont {K.~A.}\ \bibnamefont {Tanaka}},\ and\ \bibinfo
  {author} {\bibfnamefont {N.~V.}\ \bibnamefont {Zamfir}},\ }\bibfield  {title}
  {\bibinfo {title} {10 {{PW}} peak power femtosecond laser pulses at
  {{ELI-NP}}},\ }\href {https://doi.org/10.1017/hpl.2022.11} {\bibfield
  {journal} {\bibinfo  {journal} {High Power Laser Sci. Eng.}\ }\textbf
  {\bibinfo {volume} {10}},\ \bibinfo {pages} {e21} (\bibinfo {year}
  {2022})}\BibitemShut {NoStop}%
\bibitem [{\citenamefont {Sun}\ \emph {et~al.}(2022)\citenamefont {Sun},
  \citenamefont {Zhao}, \citenamefont {Xue}, \citenamefont {Lu}, \citenamefont
  {Ji}, \citenamefont {Wan}, \citenamefont {Wang}, \citenamefont {Salamin},\
  and\ \citenamefont {Li}}]{sun2022Production}%
  \BibitemOpen
  \bibfield  {author} {\bibinfo {author} {\bibfnamefont {T.}~\bibnamefont
  {Sun}}, \bibinfo {author} {\bibfnamefont {Q.}~\bibnamefont {Zhao}}, \bibinfo
  {author} {\bibfnamefont {K.}~\bibnamefont {Xue}}, \bibinfo {author}
  {\bibfnamefont {Z.-W.}\ \bibnamefont {Lu}}, \bibinfo {author} {\bibfnamefont
  {L.-L.}\ \bibnamefont {Ji}}, \bibinfo {author} {\bibfnamefont
  {F.}~\bibnamefont {Wan}}, \bibinfo {author} {\bibfnamefont {Y.}~\bibnamefont
  {Wang}}, \bibinfo {author} {\bibfnamefont {Y.~I.}\ \bibnamefont {Salamin}},\
  and\ \bibinfo {author} {\bibfnamefont {J.-X.}\ \bibnamefont {Li}},\
  }\bibfield  {title} {\bibinfo {title} {Production of polarized particle beams
  via ultraintense laser pulses},\ }\href
  {https://doi.org/10.1007/s41614-022-00099-9} {\bibfield  {journal} {\bibinfo
  {journal} {Rev. Mod. Plasma Phys.}\ }\textbf {\bibinfo {volume} {6}},\
  \bibinfo {pages} {38} (\bibinfo {year} {2022})}\BibitemShut {NoStop}%
\bibitem [{\citenamefont {Fedotov}\ \emph {et~al.}(2023)\citenamefont
  {Fedotov}, \citenamefont {Ilderton}, \citenamefont {Karbstein}, \citenamefont
  {King}, \citenamefont {Seipt}, \citenamefont {Taya},\ and\ \citenamefont
  {Torgrimsson}}]{fedotov2023Advances}%
  \BibitemOpen
  \bibfield  {author} {\bibinfo {author} {\bibfnamefont {A.}~\bibnamefont
  {Fedotov}}, \bibinfo {author} {\bibfnamefont {A.}~\bibnamefont {Ilderton}},
  \bibinfo {author} {\bibfnamefont {F.}~\bibnamefont {Karbstein}}, \bibinfo
  {author} {\bibfnamefont {B.}~\bibnamefont {King}}, \bibinfo {author}
  {\bibfnamefont {D.}~\bibnamefont {Seipt}}, \bibinfo {author} {\bibfnamefont
  {H.}~\bibnamefont {Taya}},\ and\ \bibinfo {author} {\bibfnamefont
  {G.}~\bibnamefont {Torgrimsson}},\ }\bibfield  {title} {\bibinfo {title}
  {Advances in {{QED}} with intense background fields},\ }\href
  {https://doi.org/10.1016/j.physrep.2023.01.003} {\bibfield  {journal}
  {\bibinfo  {journal} {Phys. Rep.}\ }\textbf {\bibinfo {volume} {1010}},\
  \bibinfo {pages} {1} (\bibinfo {year} {2023})}\BibitemShut {NoStop}%
\bibitem [{\citenamefont {Yu}\ \emph {et~al.}(2024)\citenamefont {Yu},
  \citenamefont {Liu}, \citenamefont {Zhao}, \citenamefont {Zhu}, \citenamefont
  {Lu}, \citenamefont {Cao}, \citenamefont {Zhang}, \citenamefont {Shao},\ and\
  \citenamefont {Sheng}}]{yu2024Bright}%
  \BibitemOpen
  \bibfield  {author} {\bibinfo {author} {\bibfnamefont {T.-P.}\ \bibnamefont
  {Yu}}, \bibinfo {author} {\bibfnamefont {K.}~\bibnamefont {Liu}}, \bibinfo
  {author} {\bibfnamefont {J.}~\bibnamefont {Zhao}}, \bibinfo {author}
  {\bibfnamefont {X.-L.}\ \bibnamefont {Zhu}}, \bibinfo {author} {\bibfnamefont
  {Y.}~\bibnamefont {Lu}}, \bibinfo {author} {\bibfnamefont {Y.}~\bibnamefont
  {Cao}}, \bibinfo {author} {\bibfnamefont {H.}~\bibnamefont {Zhang}}, \bibinfo
  {author} {\bibfnamefont {F.-Q.}\ \bibnamefont {Shao}},\ and\ \bibinfo
  {author} {\bibfnamefont {Z.-M.}\ \bibnamefont {Sheng}},\ }\bibfield  {title}
  {\bibinfo {title} {Bright {{X}}/$\gamma$-ray emission and lepton pair
  production by strong laser fields: A review},\ }\href
  {https://doi.org/10.1007/s41614-024-00158-3} {\bibfield  {journal} {\bibinfo
  {journal} {Rev. Mod. Plasma Phys.}\ }\textbf {\bibinfo {volume} {8}},\
  \bibinfo {pages} {24} (\bibinfo {year} {2024})}\BibitemShut {NoStop}%
\bibitem [{\citenamefont {Ridgers}\ \emph {et~al.}(2012)\citenamefont
  {Ridgers}, \citenamefont {Brady}, \citenamefont {Duclous}, \citenamefont
  {Kirk}, \citenamefont {Bennett}, \citenamefont {Arber}, \citenamefont
  {Robinson},\ and\ \citenamefont {Bell}}]{ridgers2012Dense}%
  \BibitemOpen
  \bibfield  {author} {\bibinfo {author} {\bibfnamefont {C.~P.}\ \bibnamefont
  {Ridgers}}, \bibinfo {author} {\bibfnamefont {C.~S.}\ \bibnamefont {Brady}},
  \bibinfo {author} {\bibfnamefont {R.}~\bibnamefont {Duclous}}, \bibinfo
  {author} {\bibfnamefont {J.~G.}\ \bibnamefont {Kirk}}, \bibinfo {author}
  {\bibfnamefont {K.}~\bibnamefont {Bennett}}, \bibinfo {author} {\bibfnamefont
  {T.~D.}\ \bibnamefont {Arber}}, \bibinfo {author} {\bibfnamefont {A.~P.~L.}\
  \bibnamefont {Robinson}},\ and\ \bibinfo {author} {\bibfnamefont {A.~R.}\
  \bibnamefont {Bell}},\ }\bibfield  {title} {\bibinfo {title} {Dense
  {{Electron-Positron Plasmas}} and {{Ultraintense}} {$\gamma$} rays from
  {{Laser-Irradiated Solids}}},\ }\href
  {https://doi.org/10.1103/PhysRevLett.108.165006} {\bibfield  {journal}
  {\bibinfo  {journal} {Phys. Rev. Lett.}\ }\textbf {\bibinfo {volume} {108}},\
  \bibinfo {pages} {165006} (\bibinfo {year} {2012})}\BibitemShut {NoStop}%
\bibitem [{\citenamefont {Zhu}\ \emph {et~al.}(2016)\citenamefont {Zhu},
  \citenamefont {Yu}, \citenamefont {Sheng}, \citenamefont {Yin}, \citenamefont
  {Turcu},\ and\ \citenamefont {Pukhov}}]{zhu2016Dense}%
  \BibitemOpen
  \bibfield  {author} {\bibinfo {author} {\bibfnamefont {X.-L.}\ \bibnamefont
  {Zhu}}, \bibinfo {author} {\bibfnamefont {T.-P.}\ \bibnamefont {Yu}},
  \bibinfo {author} {\bibfnamefont {Z.-M.}\ \bibnamefont {Sheng}}, \bibinfo
  {author} {\bibfnamefont {Y.}~\bibnamefont {Yin}}, \bibinfo {author}
  {\bibfnamefont {I.~C.~E.}\ \bibnamefont {Turcu}},\ and\ \bibinfo {author}
  {\bibfnamefont {A.}~\bibnamefont {Pukhov}},\ }\bibfield  {title} {\bibinfo
  {title} {Dense {{GeV}} electron--positron pairs generated by lasers in
  near-critical-density plasmas},\ }\href {https://doi.org/10.1038/ncomms13686}
  {\bibfield  {journal} {\bibinfo  {journal} {Nat. Commun.}\ }\textbf {\bibinfo
  {volume} {7}},\ \bibinfo {pages} {13686} (\bibinfo {year}
  {2016})}\BibitemShut {NoStop}%
\bibitem [{\citenamefont {Song}\ \emph {et~al.}(2022)\citenamefont {Song},
  \citenamefont {Wang},\ and\ \citenamefont {Li}}]{song2022Dense}%
  \BibitemOpen
  \bibfield  {author} {\bibinfo {author} {\bibfnamefont {H.-H.}\ \bibnamefont
  {Song}}, \bibinfo {author} {\bibfnamefont {W.-M.}\ \bibnamefont {Wang}},\
  and\ \bibinfo {author} {\bibfnamefont {Y.-T.}\ \bibnamefont {Li}},\
  }\bibfield  {title} {\bibinfo {title} {Dense polarized positrons from
  laser-irradiated foil targets in the qed regime},\ }\href
  {https://doi.org/10.1103/PhysRevLett.129.035001} {\bibfield  {journal}
  {\bibinfo  {journal} {Phys. Rev. Lett.}\ }\textbf {\bibinfo {volume} {129}},\
  \bibinfo {pages} {035001} (\bibinfo {year} {2022})}\BibitemShut {NoStop}%
\bibitem [{\citenamefont {Xue}\ \emph {et~al.}(2023)\citenamefont {Xue},
  \citenamefont {Sun}, \citenamefont {Wei}, \citenamefont {Li}, \citenamefont
  {Zhao}, \citenamefont {Wan}, \citenamefont {Lv}, \citenamefont {Zhao},
  \citenamefont {Xu},\ and\ \citenamefont {Li}}]{xue2023Generation}%
  \BibitemOpen
  \bibfield  {author} {\bibinfo {author} {\bibfnamefont {K.}~\bibnamefont
  {Xue}}, \bibinfo {author} {\bibfnamefont {T.}~\bibnamefont {Sun}}, \bibinfo
  {author} {\bibfnamefont {K.-J.}\ \bibnamefont {Wei}}, \bibinfo {author}
  {\bibfnamefont {Z.-P.}\ \bibnamefont {Li}}, \bibinfo {author} {\bibfnamefont
  {Q.}~\bibnamefont {Zhao}}, \bibinfo {author} {\bibfnamefont {F.}~\bibnamefont
  {Wan}}, \bibinfo {author} {\bibfnamefont {C.}~\bibnamefont {Lv}}, \bibinfo
  {author} {\bibfnamefont {Y.-T.}\ \bibnamefont {Zhao}}, \bibinfo {author}
  {\bibfnamefont {Z.-F.}\ \bibnamefont {Xu}},\ and\ \bibinfo {author}
  {\bibfnamefont {J.-X.}\ \bibnamefont {Li}},\ }\bibfield  {title} {\bibinfo
  {title} {Generation of {{High-Density High-Polarization Positrons}} via
  {{Single-Shot Strong Laser-Foil Interaction}}},\ }\href
  {https://doi.org/10.1103/PhysRevLett.131.175101} {\bibfield  {journal}
  {\bibinfo  {journal} {Phys. Rev. Lett.}\ }\textbf {\bibinfo {volume} {131}},\
  \bibinfo {pages} {175101} (\bibinfo {year} {2023})}\BibitemShut {NoStop}%
\bibitem [{\citenamefont {He}\ \emph {et~al.}(2021{\natexlab{a}})\citenamefont
  {He}, \citenamefont {Blackburn}, \citenamefont {Toncian},\ and\ \citenamefont
  {Arefiev}}]{he2021Dominance}%
  \BibitemOpen
  \bibfield  {author} {\bibinfo {author} {\bibfnamefont {Y.}~\bibnamefont
  {He}}, \bibinfo {author} {\bibfnamefont {T.~G.}\ \bibnamefont {Blackburn}},
  \bibinfo {author} {\bibfnamefont {T.}~\bibnamefont {Toncian}},\ and\ \bibinfo
  {author} {\bibfnamefont {A.~V.}\ \bibnamefont {Arefiev}},\ }\bibfield
  {title} {\bibinfo {title} {Dominance of {$\gamma$}-{$\gamma$}
  electron-positron pair creation in a plasma driven by high-intensity
  lasers},\ }\href {https://doi.org/10.1038/s42005-021-00636-x} {\bibfield
  {journal} {\bibinfo  {journal} {Commun. Phys.}\ }\textbf {\bibinfo {volume}
  {4}},\ \bibinfo {pages} {139} (\bibinfo {year}
  {2021}{\natexlab{a}})}\BibitemShut {NoStop}%
\bibitem [{\citenamefont {Sugimoto}\ \emph {et~al.}(2023)\citenamefont
  {Sugimoto}, \citenamefont {He}, \citenamefont {Iwata}, \citenamefont {Yeh},
  \citenamefont {Tangtartharakul}, \citenamefont {Arefiev},\ and\ \citenamefont
  {Sentoku}}]{sugimoto2023Positron}%
  \BibitemOpen
  \bibfield  {author} {\bibinfo {author} {\bibfnamefont {K.}~\bibnamefont
  {Sugimoto}}, \bibinfo {author} {\bibfnamefont {Y.}~\bibnamefont {He}},
  \bibinfo {author} {\bibfnamefont {N.}~\bibnamefont {Iwata}}, \bibinfo
  {author} {\bibfnamefont {I.-L.}\ \bibnamefont {Yeh}}, \bibinfo {author}
  {\bibfnamefont {K.}~\bibnamefont {Tangtartharakul}}, \bibinfo {author}
  {\bibfnamefont {A.}~\bibnamefont {Arefiev}},\ and\ \bibinfo {author}
  {\bibfnamefont {Y.}~\bibnamefont {Sentoku}},\ }\bibfield  {title} {\bibinfo
  {title} {Positron {{Generation}} and {{Acceleration}} in a {{Self-Organized
  Photon Collider Enabled}} by an {{Ultraintense Laser Pulse}}},\ }\href
  {https://doi.org/10.1103/PhysRevLett.131.065102} {\bibfield  {journal}
  {\bibinfo  {journal} {Phys. Rev. Lett.}\ }\textbf {\bibinfo {volume} {131}},\
  \bibinfo {pages} {065102} (\bibinfo {year} {2023})}\BibitemShut {NoStop}%
\bibitem [{\citenamefont {Song}\ \emph {et~al.}(2024)\citenamefont {Song},
  \citenamefont {Wang}, \citenamefont {Chen},\ and\ \citenamefont
  {Sheng}}]{song2024From}%
  \BibitemOpen
  \bibfield  {author} {\bibinfo {author} {\bibfnamefont {H.-H.}\ \bibnamefont
  {Song}}, \bibinfo {author} {\bibfnamefont {W.-M.}\ \bibnamefont {Wang}},
  \bibinfo {author} {\bibfnamefont {M.}~\bibnamefont {Chen}},\ and\ \bibinfo
  {author} {\bibfnamefont {Z.-M.}\ \bibnamefont {Sheng}},\ }\bibfield  {title}
  {\bibinfo {title} {From linear to nonlinear {{Breit-Wheeler}} pair production
  in laser-solid interactions},\ }\href
  {https://doi.org/10.1103/PhysRevE.109.035204} {\bibfield  {journal} {\bibinfo
   {journal} {Phys. Rev. E}\ }\textbf {\bibinfo {volume} {109}},\ \bibinfo
  {pages} {035204} (\bibinfo {year} {2024})}\BibitemShut {NoStop}%
\bibitem [{\citenamefont {Zhao}\ \emph {et~al.}(2023)\citenamefont {Zhao},
  \citenamefont {Sun}, \citenamefont {Xue}, \citenamefont {Wan},\ and\
  \citenamefont {Li}}]{zhao2023Cascade}%
  \BibitemOpen
  \bibfield  {author} {\bibinfo {author} {\bibfnamefont {Q.}~\bibnamefont
  {Zhao}}, \bibinfo {author} {\bibfnamefont {T.}~\bibnamefont {Sun}}, \bibinfo
  {author} {\bibfnamefont {K.}~\bibnamefont {Xue}}, \bibinfo {author}
  {\bibfnamefont {F.}~\bibnamefont {Wan}},\ and\ \bibinfo {author}
  {\bibfnamefont {J.-X.}\ \bibnamefont {Li}},\ }\bibfield  {title} {\bibinfo
  {title} {Cascade of polarized {{Compton}} scattering and {{Breit-Wheeler}}
  pair production},\ }\href {https://doi.org/10.1103/PhysRevD.108.116012}
  {\bibfield  {journal} {\bibinfo  {journal} {Phys. Rev. D}\ }\textbf {\bibinfo
  {volume} {108}},\ \bibinfo {pages} {116012} (\bibinfo {year}
  {2023})}\BibitemShut {NoStop}%
\bibitem [{\citenamefont {Dou}\ \emph {et~al.}(2026)\citenamefont {Dou},
  \citenamefont {Zhao}, \citenamefont {Wan}, \citenamefont {Lv}, \citenamefont
  {Guo},\ and\ \citenamefont {Li}}]{dou2026Generation}%
  \BibitemOpen
  \bibfield  {author} {\bibinfo {author} {\bibfnamefont {Z.-K.}\ \bibnamefont
  {Dou}}, \bibinfo {author} {\bibfnamefont {Q.}~\bibnamefont {Zhao}}, \bibinfo
  {author} {\bibfnamefont {F.}~\bibnamefont {Wan}}, \bibinfo {author}
  {\bibfnamefont {C.}~\bibnamefont {Lv}}, \bibinfo {author} {\bibfnamefont
  {B.}~\bibnamefont {Guo}},\ and\ \bibinfo {author} {\bibfnamefont {J.-X.}\
  \bibnamefont {Li}},\ }\href {https://arxiv.org/abs/2603.26383} {\bibinfo
  {title} {Generation of polarized overdense pair-photon fireball via
  laser-driven nonlinear-linear qed cascade}} (\bibinfo {year} {2026}),\
  \Eprint {https://arxiv.org/abs/2603.26383} {arXiv:2603.26383
  [physics.plasm-ph]} \BibitemShut {NoStop}%
\bibitem [{\citenamefont {Corde}\ \emph {et~al.}(2015)\citenamefont {Corde},
  \citenamefont {Adli}, \citenamefont {Allen}, \citenamefont {An},
  \citenamefont {Clarke}, \citenamefont {Clayton}, \citenamefont {Delahaye},
  \citenamefont {Frederico}, \citenamefont {Gessner}, \citenamefont {Green},
  \citenamefont {Hogan}, \citenamefont {Joshi}, \citenamefont {Lipkowitz},
  \citenamefont {Litos}, \citenamefont {Lu}, \citenamefont {Marsh},
  \citenamefont {Mori}, \citenamefont {Schmeltz}, \citenamefont
  {{Vafaei-Najafabadi}}, \citenamefont {Walz}, \citenamefont {Yakimenko},\ and\
  \citenamefont {Yocky}}]{corde2015Multi}%
  \BibitemOpen
  \bibfield  {author} {\bibinfo {author} {\bibfnamefont {S.}~\bibnamefont
  {Corde}}, \bibinfo {author} {\bibfnamefont {E.}~\bibnamefont {Adli}},
  \bibinfo {author} {\bibfnamefont {J.~M.}\ \bibnamefont {Allen}}, \bibinfo
  {author} {\bibfnamefont {W.}~\bibnamefont {An}}, \bibinfo {author}
  {\bibfnamefont {C.~I.}\ \bibnamefont {Clarke}}, \bibinfo {author}
  {\bibfnamefont {C.~E.}\ \bibnamefont {Clayton}}, \bibinfo {author}
  {\bibfnamefont {J.~P.}\ \bibnamefont {Delahaye}}, \bibinfo {author}
  {\bibfnamefont {J.}~\bibnamefont {Frederico}}, \bibinfo {author}
  {\bibfnamefont {S.}~\bibnamefont {Gessner}}, \bibinfo {author} {\bibfnamefont
  {S.~Z.}\ \bibnamefont {Green}}, \bibinfo {author} {\bibfnamefont {M.~J.}\
  \bibnamefont {Hogan}}, \bibinfo {author} {\bibfnamefont {C.}~\bibnamefont
  {Joshi}}, \bibinfo {author} {\bibfnamefont {N.}~\bibnamefont {Lipkowitz}},
  \bibinfo {author} {\bibfnamefont {M.}~\bibnamefont {Litos}}, \bibinfo
  {author} {\bibfnamefont {W.}~\bibnamefont {Lu}}, \bibinfo {author}
  {\bibfnamefont {K.~A.}\ \bibnamefont {Marsh}}, \bibinfo {author}
  {\bibfnamefont {W.~B.}\ \bibnamefont {Mori}}, \bibinfo {author}
  {\bibfnamefont {M.}~\bibnamefont {Schmeltz}}, \bibinfo {author}
  {\bibfnamefont {N.}~\bibnamefont {{Vafaei-Najafabadi}}}, \bibinfo {author}
  {\bibfnamefont {D.}~\bibnamefont {Walz}}, \bibinfo {author} {\bibfnamefont
  {V.}~\bibnamefont {Yakimenko}},\ and\ \bibinfo {author} {\bibfnamefont
  {G.}~\bibnamefont {Yocky}},\ }\bibfield  {title} {\bibinfo {title}
  {Multi-gigaelectronvolt acceleration of positrons in a self-loaded plasma
  wakefield},\ }\href {https://doi.org/10.1038/nature14890} {\bibfield
  {journal} {\bibinfo  {journal} {Nature}\ }\textbf {\bibinfo {volume} {524}},\
  \bibinfo {pages} {442} (\bibinfo {year} {2015})}\BibitemShut {NoStop}%
\bibitem [{\citenamefont {Doche}\ \emph {et~al.}(2017)\citenamefont {Doche},
  \citenamefont {Beekman}, \citenamefont {Corde}, \citenamefont {Allen},
  \citenamefont {Clarke}, \citenamefont {Frederico}, \citenamefont {Gessner},
  \citenamefont {Green}, \citenamefont {Hogan}, \citenamefont {O'Shea},
  \citenamefont {Yakimenko}, \citenamefont {An}, \citenamefont {Clayton},
  \citenamefont {Joshi}, \citenamefont {Marsh}, \citenamefont {Mori},
  \citenamefont {{Vafaei-Najafabadi}}, \citenamefont {Litos}, \citenamefont
  {Adli}, \citenamefont {Lindstr{\o}m},\ and\ \citenamefont
  {Lu}}]{doche2017Acceleration}%
  \BibitemOpen
  \bibfield  {author} {\bibinfo {author} {\bibfnamefont {A.}~\bibnamefont
  {Doche}}, \bibinfo {author} {\bibfnamefont {C.}~\bibnamefont {Beekman}},
  \bibinfo {author} {\bibfnamefont {S.}~\bibnamefont {Corde}}, \bibinfo
  {author} {\bibfnamefont {J.~M.}\ \bibnamefont {Allen}}, \bibinfo {author}
  {\bibfnamefont {C.~I.}\ \bibnamefont {Clarke}}, \bibinfo {author}
  {\bibfnamefont {J.}~\bibnamefont {Frederico}}, \bibinfo {author}
  {\bibfnamefont {S.~J.}\ \bibnamefont {Gessner}}, \bibinfo {author}
  {\bibfnamefont {S.~Z.}\ \bibnamefont {Green}}, \bibinfo {author}
  {\bibfnamefont {M.~J.}\ \bibnamefont {Hogan}}, \bibinfo {author}
  {\bibfnamefont {B.}~\bibnamefont {O'Shea}}, \bibinfo {author} {\bibfnamefont
  {V.}~\bibnamefont {Yakimenko}}, \bibinfo {author} {\bibfnamefont
  {W.}~\bibnamefont {An}}, \bibinfo {author} {\bibfnamefont {C.~E.}\
  \bibnamefont {Clayton}}, \bibinfo {author} {\bibfnamefont {C.}~\bibnamefont
  {Joshi}}, \bibinfo {author} {\bibfnamefont {K.~A.}\ \bibnamefont {Marsh}},
  \bibinfo {author} {\bibfnamefont {W.~B.}\ \bibnamefont {Mori}}, \bibinfo
  {author} {\bibfnamefont {N.}~\bibnamefont {{Vafaei-Najafabadi}}}, \bibinfo
  {author} {\bibfnamefont {M.~D.}\ \bibnamefont {Litos}}, \bibinfo {author}
  {\bibfnamefont {E.}~\bibnamefont {Adli}}, \bibinfo {author} {\bibfnamefont
  {C.~A.}\ \bibnamefont {Lindstr{\o}m}},\ and\ \bibinfo {author} {\bibfnamefont
  {W.}~\bibnamefont {Lu}},\ }\bibfield  {title} {\bibinfo {title} {Acceleration
  of a trailing positron bunch in a plasma wakefield accelerator},\ }\href
  {https://doi.org/10.1038/s41598-017-14524-4} {\bibfield  {journal} {\bibinfo
  {journal} {Sci. Rep.}\ }\textbf {\bibinfo {volume} {7}},\ \bibinfo {pages}
  {14180} (\bibinfo {year} {2017})}\BibitemShut {NoStop}%
\bibitem [{\citenamefont {Silva}\ and\ \citenamefont
  {Vieira}(2023)}]{silva2023Positron}%
  \BibitemOpen
  \bibfield  {author} {\bibinfo {author} {\bibfnamefont {T.}~\bibnamefont
  {Silva}}\ and\ \bibinfo {author} {\bibfnamefont {J.}~\bibnamefont {Vieira}},\
  }\bibfield  {title} {\bibinfo {title} {Positron acceleration in plasma waves
  driven by non-neutral fireball beams},\ }\href
  {https://doi.org/10.1103/PhysRevAccelBeams.26.091301} {\bibfield  {journal}
  {\bibinfo  {journal} {Phys. Rev. Accel. Beams}\ }\textbf {\bibinfo {volume}
  {26}},\ \bibinfo {pages} {091301} (\bibinfo {year} {2023})}\BibitemShut
  {NoStop}%
\bibitem [{\citenamefont {Gessner}\ \emph {et~al.}(2016)\citenamefont
  {Gessner}, \citenamefont {Adli}, \citenamefont {Allen}, \citenamefont {An},
  \citenamefont {Clarke}, \citenamefont {Clayton}, \citenamefont {Corde},
  \citenamefont {Delahaye}, \citenamefont {Frederico}, \citenamefont {Green},
  \citenamefont {Hast}, \citenamefont {Hogan}, \citenamefont {Joshi},
  \citenamefont {Lindstr{\o}m}, \citenamefont {Lipkowitz}, \citenamefont
  {Litos}, \citenamefont {Lu}, \citenamefont {Marsh}, \citenamefont {Mori},
  \citenamefont {O'Shea}, \citenamefont {{Vafaei-Najafabadi}}, \citenamefont
  {Walz}, \citenamefont {Yakimenko},\ and\ \citenamefont
  {Yocky}}]{gessner2016Demonstration}%
  \BibitemOpen
  \bibfield  {author} {\bibinfo {author} {\bibfnamefont {S.}~\bibnamefont
  {Gessner}}, \bibinfo {author} {\bibfnamefont {E.}~\bibnamefont {Adli}},
  \bibinfo {author} {\bibfnamefont {J.~M.}\ \bibnamefont {Allen}}, \bibinfo
  {author} {\bibfnamefont {W.}~\bibnamefont {An}}, \bibinfo {author}
  {\bibfnamefont {C.~I.}\ \bibnamefont {Clarke}}, \bibinfo {author}
  {\bibfnamefont {C.~E.}\ \bibnamefont {Clayton}}, \bibinfo {author}
  {\bibfnamefont {S.}~\bibnamefont {Corde}}, \bibinfo {author} {\bibfnamefont
  {J.~P.}\ \bibnamefont {Delahaye}}, \bibinfo {author} {\bibfnamefont
  {J.}~\bibnamefont {Frederico}}, \bibinfo {author} {\bibfnamefont {S.~Z.}\
  \bibnamefont {Green}}, \bibinfo {author} {\bibfnamefont {C.}~\bibnamefont
  {Hast}}, \bibinfo {author} {\bibfnamefont {M.~J.}\ \bibnamefont {Hogan}},
  \bibinfo {author} {\bibfnamefont {C.}~\bibnamefont {Joshi}}, \bibinfo
  {author} {\bibfnamefont {C.~A.}\ \bibnamefont {Lindstr{\o}m}}, \bibinfo
  {author} {\bibfnamefont {N.}~\bibnamefont {Lipkowitz}}, \bibinfo {author}
  {\bibfnamefont {M.}~\bibnamefont {Litos}}, \bibinfo {author} {\bibfnamefont
  {W.}~\bibnamefont {Lu}}, \bibinfo {author} {\bibfnamefont {K.~A.}\
  \bibnamefont {Marsh}}, \bibinfo {author} {\bibfnamefont {W.~B.}\ \bibnamefont
  {Mori}}, \bibinfo {author} {\bibfnamefont {B.}~\bibnamefont {O'Shea}},
  \bibinfo {author} {\bibfnamefont {N.}~\bibnamefont {{Vafaei-Najafabadi}}},
  \bibinfo {author} {\bibfnamefont {D.}~\bibnamefont {Walz}}, \bibinfo {author}
  {\bibfnamefont {V.}~\bibnamefont {Yakimenko}},\ and\ \bibinfo {author}
  {\bibfnamefont {G.}~\bibnamefont {Yocky}},\ }\bibfield  {title} {\bibinfo
  {title} {Demonstration of a positron beam-driven hollow channel plasma
  wakefield accelerator},\ }\href {https://doi.org/10.1038/ncomms11785}
  {\bibfield  {journal} {\bibinfo  {journal} {Nat. Commun.}\ }\textbf {\bibinfo
  {volume} {7}},\ \bibinfo {pages} {11785} (\bibinfo {year}
  {2016})}\BibitemShut {NoStop}%
\bibitem [{\citenamefont {Silva}\ \emph {et~al.}(2021)\citenamefont {Silva},
  \citenamefont {Amorim}, \citenamefont {Downer}, \citenamefont {Hogan},
  \citenamefont {Yakimenko}, \citenamefont {Zgadzaj},\ and\ \citenamefont
  {Vieira}}]{silva2021Stable}%
  \BibitemOpen
  \bibfield  {author} {\bibinfo {author} {\bibfnamefont {T.}~\bibnamefont
  {Silva}}, \bibinfo {author} {\bibfnamefont {L.~D.}\ \bibnamefont {Amorim}},
  \bibinfo {author} {\bibfnamefont {M.~C.}\ \bibnamefont {Downer}}, \bibinfo
  {author} {\bibfnamefont {M.~J.}\ \bibnamefont {Hogan}}, \bibinfo {author}
  {\bibfnamefont {V.}~\bibnamefont {Yakimenko}}, \bibinfo {author}
  {\bibfnamefont {R.}~\bibnamefont {Zgadzaj}},\ and\ \bibinfo {author}
  {\bibfnamefont {J.}~\bibnamefont {Vieira}},\ }\bibfield  {title} {\bibinfo
  {title} {Stable positron acceleration in thin, warm, hollow plasma
  channels},\ }\href {https://doi.org/10.1103/PhysRevLett.127.104801}
  {\bibfield  {journal} {\bibinfo  {journal} {Phys. Rev. Lett.}\ }\textbf
  {\bibinfo {volume} {127}},\ \bibinfo {pages} {104801} (\bibinfo {year}
  {2021})}\BibitemShut {NoStop}%
\bibitem [{\citenamefont {Zhou}\ \emph {et~al.}(2021)\citenamefont {Zhou},
  \citenamefont {Hua}, \citenamefont {An}, \citenamefont {Mori}, \citenamefont
  {Joshi}, \citenamefont {Gao},\ and\ \citenamefont {Lu}}]{zhou2021High}%
  \BibitemOpen
  \bibfield  {author} {\bibinfo {author} {\bibfnamefont {S.}~\bibnamefont
  {Zhou}}, \bibinfo {author} {\bibfnamefont {J.}~\bibnamefont {Hua}}, \bibinfo
  {author} {\bibfnamefont {W.}~\bibnamefont {An}}, \bibinfo {author}
  {\bibfnamefont {W.~B.}\ \bibnamefont {Mori}}, \bibinfo {author}
  {\bibfnamefont {C.}~\bibnamefont {Joshi}}, \bibinfo {author} {\bibfnamefont
  {J.}~\bibnamefont {Gao}},\ and\ \bibinfo {author} {\bibfnamefont
  {W.}~\bibnamefont {Lu}},\ }\bibfield  {title} {\bibinfo {title} {High
  efficiency uniform wakefield acceleration of a positron beam using stable
  asymmetric mode in a hollow channel plasma},\ }\href
  {https://doi.org/10.1103/PhysRevLett.127.174801} {\bibfield  {journal}
  {\bibinfo  {journal} {Phys. Rev. Lett.}\ }\textbf {\bibinfo {volume} {127}},\
  \bibinfo {pages} {174801} (\bibinfo {year} {2021})}\BibitemShut {NoStop}%
\bibitem [{\citenamefont {Vieira}\ and\ \citenamefont
  {Mendon\ifmmode~\mbox{\c{c}}\else \c{c}\fi{}a}(2014)}]{vieira2014Nonlinear}%
  \BibitemOpen
  \bibfield  {author} {\bibinfo {author} {\bibfnamefont {J.}~\bibnamefont
  {Vieira}}\ and\ \bibinfo {author} {\bibfnamefont {J.~T.}\ \bibnamefont
  {Mendon\ifmmode~\mbox{\c{c}}\else \c{c}\fi{}a}},\ }\bibfield  {title}
  {\bibinfo {title} {Nonlinear laser driven donut wakefields for positron and
  electron acceleration},\ }\href
  {https://doi.org/10.1103/PhysRevLett.112.215001} {\bibfield  {journal}
  {\bibinfo  {journal} {Phys. Rev. Lett.}\ }\textbf {\bibinfo {volume} {112}},\
  \bibinfo {pages} {215001} (\bibinfo {year} {2014})}\BibitemShut {NoStop}%
\bibitem [{\citenamefont {Jain}\ \emph {et~al.}(2015)\citenamefont {Jain},
  \citenamefont {Antonsen},\ and\ \citenamefont {Palastro}}]{jain2015Positron}%
  \BibitemOpen
  \bibfield  {author} {\bibinfo {author} {\bibfnamefont {N.}~\bibnamefont
  {Jain}}, \bibinfo {author} {\bibfnamefont {T.~M.}\ \bibnamefont {Antonsen}},\
  and\ \bibinfo {author} {\bibfnamefont {J.~P.}\ \bibnamefont {Palastro}},\
  }\bibfield  {title} {\bibinfo {title} {Positron acceleration by plasma
  wakefields driven by a hollow electron beam},\ }\href
  {https://doi.org/10.1103/PhysRevLett.115.195001} {\bibfield  {journal}
  {\bibinfo  {journal} {Phys. Rev. Lett.}\ }\textbf {\bibinfo {volume} {115}},\
  \bibinfo {pages} {195001} (\bibinfo {year} {2015})}\BibitemShut {NoStop}%
\bibitem [{\citenamefont {Liu}\ \emph {et~al.}(2022)\citenamefont {Liu},
  \citenamefont {Xue}, \citenamefont {Wan}, \citenamefont {Chen}, \citenamefont
  {Li}, \citenamefont {Liu}, \citenamefont {Weng}, \citenamefont {Sheng},\ and\
  \citenamefont {Zhang}}]{liu2022Trapping}%
  \BibitemOpen
  \bibfield  {author} {\bibinfo {author} {\bibfnamefont {W.-Y.}\ \bibnamefont
  {Liu}}, \bibinfo {author} {\bibfnamefont {K.}~\bibnamefont {Xue}}, \bibinfo
  {author} {\bibfnamefont {F.}~\bibnamefont {Wan}}, \bibinfo {author}
  {\bibfnamefont {M.}~\bibnamefont {Chen}}, \bibinfo {author} {\bibfnamefont
  {J.-X.}\ \bibnamefont {Li}}, \bibinfo {author} {\bibfnamefont
  {F.}~\bibnamefont {Liu}}, \bibinfo {author} {\bibfnamefont {S.-M.}\
  \bibnamefont {Weng}}, \bibinfo {author} {\bibfnamefont {Z.-M.}\ \bibnamefont
  {Sheng}},\ and\ \bibinfo {author} {\bibfnamefont {J.}~\bibnamefont {Zhang}},\
  }\bibfield  {title} {\bibinfo {title} {Trapping and acceleration of
  spin-polarized positrons from {$\gamma$} photon splitting in wakefields},\
  }\href {https://doi.org/10.1103/PhysRevResearch.4.L022028} {\bibfield
  {journal} {\bibinfo  {journal} {Phys. Rev. Res.}\ }\textbf {\bibinfo {volume}
  {4}},\ \bibinfo {pages} {L022028} (\bibinfo {year} {2022})}\BibitemShut
  {NoStop}%
\bibitem [{\citenamefont {Sun}\ \emph {et~al.}(2025)\citenamefont {Sun},
  \citenamefont {Dou}, \citenamefont {Huang}, \citenamefont {Wan},
  \citenamefont {Zhao},\ and\ \citenamefont {Li}}]{sun2025Generation}%
  \BibitemOpen
  \bibfield  {author} {\bibinfo {author} {\bibfnamefont {T.}~\bibnamefont
  {Sun}}, \bibinfo {author} {\bibfnamefont {Z.-K.}\ \bibnamefont {Dou}},
  \bibinfo {author} {\bibfnamefont {Y.-Q.}\ \bibnamefont {Huang}}, \bibinfo
  {author} {\bibfnamefont {F.}~\bibnamefont {Wan}}, \bibinfo {author}
  {\bibfnamefont {Q.}~\bibnamefont {Zhao}},\ and\ \bibinfo {author}
  {\bibfnamefont {J.-X.}\ \bibnamefont {Li}},\ }\href
  {https://doi.org/10.48550/arXiv.2508.11148} {\bibinfo {title} {Generation of
  {{Ultrabrilliant Positron Beam}} via {{Superponderomotive Injection}} in
  {{Laser Wakefield Acceleration}}}} (\bibinfo {year} {2025}),\ \Eprint
  {https://arxiv.org/abs/2508.11148} {arXiv:2508.11148 [physics]} \BibitemShut
  {NoStop}%
\bibitem [{\citenamefont {Wan}\ \emph {et~al.}(2023)\citenamefont {Wan},
  \citenamefont {Lv}, \citenamefont {Xue}, \citenamefont {Dou}, \citenamefont
  {Zhao}, \citenamefont {Ababekri}, \citenamefont {Wei}, \citenamefont {Li},
  \citenamefont {Zhao},\ and\ \citenamefont {Li}}]{Wan2023Simulations}%
  \BibitemOpen
  \bibfield  {author} {\bibinfo {author} {\bibfnamefont {F.}~\bibnamefont
  {Wan}}, \bibinfo {author} {\bibfnamefont {C.}~\bibnamefont {Lv}}, \bibinfo
  {author} {\bibfnamefont {K.}~\bibnamefont {Xue}}, \bibinfo {author}
  {\bibfnamefont {Z.-K.}\ \bibnamefont {Dou}}, \bibinfo {author} {\bibfnamefont
  {Q.}~\bibnamefont {Zhao}}, \bibinfo {author} {\bibfnamefont {M.}~\bibnamefont
  {Ababekri}}, \bibinfo {author} {\bibfnamefont {W.-Q.}\ \bibnamefont {Wei}},
  \bibinfo {author} {\bibfnamefont {Z.-P.}\ \bibnamefont {Li}}, \bibinfo
  {author} {\bibfnamefont {Y.-T.}\ \bibnamefont {Zhao}},\ and\ \bibinfo
  {author} {\bibfnamefont {J.-X.}\ \bibnamefont {Li}},\ }\bibfield  {title}
  {\bibinfo {title} {{Simulations of spin/polarization-resolved laser–plasma
  interactions in the nonlinear QED regime}},\ }\href
  {https://doi.org/10.1063/5.0163929} {\bibfield  {journal} {\bibinfo
  {journal} {Matter Radiat. Extremes}\ }\textbf {\bibinfo {volume} {8}},\
  \bibinfo {pages} {064002} (\bibinfo {year} {2023})}\BibitemShut {NoStop}%
\bibitem [{\citenamefont {Yu}\ \emph {et~al.}(2019)\citenamefont {Yu},
  \citenamefont {Lu}, \citenamefont {Takahashi}, \citenamefont {Hu},
  \citenamefont {Gong}, \citenamefont {Ma}, \citenamefont {Huang},
  \citenamefont {Chen},\ and\ \citenamefont {Yan}}]{yu2019Creation}%
  \BibitemOpen
  \bibfield  {author} {\bibinfo {author} {\bibfnamefont {J.~Q.}\ \bibnamefont
  {Yu}}, \bibinfo {author} {\bibfnamefont {H.~Y.}\ \bibnamefont {Lu}}, \bibinfo
  {author} {\bibfnamefont {T.}~\bibnamefont {Takahashi}}, \bibinfo {author}
  {\bibfnamefont {R.~H.}\ \bibnamefont {Hu}}, \bibinfo {author} {\bibfnamefont
  {Z.}~\bibnamefont {Gong}}, \bibinfo {author} {\bibfnamefont {W.~J.}\
  \bibnamefont {Ma}}, \bibinfo {author} {\bibfnamefont {Y.~S.}\ \bibnamefont
  {Huang}}, \bibinfo {author} {\bibfnamefont {C.~E.}\ \bibnamefont {Chen}},\
  and\ \bibinfo {author} {\bibfnamefont {X.~Q.}\ \bibnamefont {Yan}},\
  }\bibfield  {title} {\bibinfo {title} {Creation of {{Electron-Positron
  Pairs}} in {{Photon-Photon Collisions Driven}} by 10-{{PW Laser Pulses}}},\
  }\href {https://doi.org/10.1103/PhysRevLett.122.014802} {\bibfield  {journal}
  {\bibinfo  {journal} {Phys. Rev. Lett.}\ }\textbf {\bibinfo {volume} {122}},\
  \bibinfo {pages} {014802} (\bibinfo {year} {2019})}\BibitemShut {NoStop}%
\bibitem [{\citenamefont {He}\ \emph {et~al.}(2021{\natexlab{b}})\citenamefont
  {He}, \citenamefont {Yeh}, \citenamefont {Blackburn},\ and\ \citenamefont
  {Arefiev}}]{he2021Single}%
  \BibitemOpen
  \bibfield  {author} {\bibinfo {author} {\bibfnamefont {Y.}~\bibnamefont
  {He}}, \bibinfo {author} {\bibfnamefont {I.-L.}\ \bibnamefont {Yeh}},
  \bibinfo {author} {\bibfnamefont {T.~G.}\ \bibnamefont {Blackburn}},\ and\
  \bibinfo {author} {\bibfnamefont {A.}~\bibnamefont {Arefiev}},\ }\bibfield
  {title} {\bibinfo {title} {A single-laser scheme for observation of linear
  {{Breit}}--{{Wheeler}} electron--positron pair creation},\ }\href
  {https://doi.org/10.1088/1367-2630/ac3049} {\bibfield  {journal} {\bibinfo
  {journal} {New J. Phys.}\ }\textbf {\bibinfo {volume} {23}},\ \bibinfo
  {pages} {115005} (\bibinfo {year} {2021}{\natexlab{b}})}\BibitemShut
  {NoStop}%
\bibitem [{\citenamefont {Kettle}\ \emph {et~al.}(2021)\citenamefont {Kettle},
  \citenamefont {Hollatz}, \citenamefont {Gerstmayr}, \citenamefont {Samarin},
  \citenamefont {Alejo}, \citenamefont {Astbury}, \citenamefont {Baird},
  \citenamefont {Bohlen}, \citenamefont {Campbell}, \citenamefont {Colgan},
  \citenamefont {Dannheim}, \citenamefont {Gregory}, \citenamefont {Harsh},
  \citenamefont {Hatfield}, \citenamefont {Hinojosa}, \citenamefont {Katzir},
  \citenamefont {Morton}, \citenamefont {Murphy}, \citenamefont {Nurnberg},
  \citenamefont {Osterhoff}, \citenamefont {{P{\'e}rez-Callejo}}, \citenamefont
  {P{\~o}der}, \citenamefont {Rajeev}, \citenamefont {Roedel}, \citenamefont
  {Roeder}, \citenamefont {Salgado}, \citenamefont {Sarri}, \citenamefont
  {Seidel}, \citenamefont {Spannagel}, \citenamefont {Spindloe}, \citenamefont
  {Steinke}, \citenamefont {Streeter}, \citenamefont {Thomas}, \citenamefont
  {Underwood}, \citenamefont {Watt}, \citenamefont {Zepf}, \citenamefont
  {Rose},\ and\ \citenamefont {Mangles}}]{kettle2021Laser}%
  \BibitemOpen
  \bibfield  {author} {\bibinfo {author} {\bibfnamefont {B.}~\bibnamefont
  {Kettle}}, \bibinfo {author} {\bibfnamefont {D.}~\bibnamefont {Hollatz}},
  \bibinfo {author} {\bibfnamefont {E.}~\bibnamefont {Gerstmayr}}, \bibinfo
  {author} {\bibfnamefont {G.~M.}\ \bibnamefont {Samarin}}, \bibinfo {author}
  {\bibfnamefont {A.}~\bibnamefont {Alejo}}, \bibinfo {author} {\bibfnamefont
  {S.}~\bibnamefont {Astbury}}, \bibinfo {author} {\bibfnamefont
  {C.}~\bibnamefont {Baird}}, \bibinfo {author} {\bibfnamefont
  {S.}~\bibnamefont {Bohlen}}, \bibinfo {author} {\bibfnamefont
  {M.}~\bibnamefont {Campbell}}, \bibinfo {author} {\bibfnamefont
  {C.}~\bibnamefont {Colgan}}, \bibinfo {author} {\bibfnamefont
  {D.}~\bibnamefont {Dannheim}}, \bibinfo {author} {\bibfnamefont
  {C.}~\bibnamefont {Gregory}}, \bibinfo {author} {\bibfnamefont
  {H.}~\bibnamefont {Harsh}}, \bibinfo {author} {\bibfnamefont
  {P.}~\bibnamefont {Hatfield}}, \bibinfo {author} {\bibfnamefont
  {J.}~\bibnamefont {Hinojosa}}, \bibinfo {author} {\bibfnamefont
  {Y.}~\bibnamefont {Katzir}}, \bibinfo {author} {\bibfnamefont
  {J.}~\bibnamefont {Morton}}, \bibinfo {author} {\bibfnamefont {C.~D.}\
  \bibnamefont {Murphy}}, \bibinfo {author} {\bibfnamefont {A.}~\bibnamefont
  {Nurnberg}}, \bibinfo {author} {\bibfnamefont {J.}~\bibnamefont {Osterhoff}},
  \bibinfo {author} {\bibfnamefont {G.}~\bibnamefont {{P{\'e}rez-Callejo}}},
  \bibinfo {author} {\bibfnamefont {K.}~\bibnamefont {P{\~o}der}}, \bibinfo
  {author} {\bibfnamefont {P.~P.}\ \bibnamefont {Rajeev}}, \bibinfo {author}
  {\bibfnamefont {C.}~\bibnamefont {Roedel}}, \bibinfo {author} {\bibfnamefont
  {F.}~\bibnamefont {Roeder}}, \bibinfo {author} {\bibfnamefont {F.~C.}\
  \bibnamefont {Salgado}}, \bibinfo {author} {\bibfnamefont {G.}~\bibnamefont
  {Sarri}}, \bibinfo {author} {\bibfnamefont {A.}~\bibnamefont {Seidel}},
  \bibinfo {author} {\bibfnamefont {S.}~\bibnamefont {Spannagel}}, \bibinfo
  {author} {\bibfnamefont {C.}~\bibnamefont {Spindloe}}, \bibinfo {author}
  {\bibfnamefont {S.}~\bibnamefont {Steinke}}, \bibinfo {author} {\bibfnamefont
  {M.~J.~V.}\ \bibnamefont {Streeter}}, \bibinfo {author} {\bibfnamefont
  {A.~G.~R.}\ \bibnamefont {Thomas}}, \bibinfo {author} {\bibfnamefont
  {C.}~\bibnamefont {Underwood}}, \bibinfo {author} {\bibfnamefont
  {R.}~\bibnamefont {Watt}}, \bibinfo {author} {\bibfnamefont {M.}~\bibnamefont
  {Zepf}}, \bibinfo {author} {\bibfnamefont {S.~J.}\ \bibnamefont {Rose}},\
  and\ \bibinfo {author} {\bibfnamefont {S.~P.~D.}\ \bibnamefont {Mangles}},\
  }\bibfield  {title} {\bibinfo {title} {A laser--plasma platform for
  photon--photon physics: The two photon {{Breit}}--{{Wheeler}} process},\
  }\href {https://doi.org/10.1088/1367-2630/ac3048} {\bibfield  {journal}
  {\bibinfo  {journal} {New J. Phys.}\ }\textbf {\bibinfo {volume} {23}},\
  \bibinfo {pages} {115006} (\bibinfo {year} {2021})}\BibitemShut {NoStop}%
\bibitem [{\citenamefont {Zhao}\ \emph {et~al.}(2022)\citenamefont {Zhao},
  \citenamefont {Tang}, \citenamefont {Wan}, \citenamefont {Liu}, \citenamefont
  {Liu}, \citenamefont {Yang}, \citenamefont {Yu}, \citenamefont {Ren},
  \citenamefont {Xu}, \citenamefont {Zhao}, \citenamefont {Huang},\ and\
  \citenamefont {Li}}]{zhao2022Signatures}%
  \BibitemOpen
  \bibfield  {author} {\bibinfo {author} {\bibfnamefont {Q.}~\bibnamefont
  {Zhao}}, \bibinfo {author} {\bibfnamefont {L.}~\bibnamefont {Tang}}, \bibinfo
  {author} {\bibfnamefont {F.}~\bibnamefont {Wan}}, \bibinfo {author}
  {\bibfnamefont {B.-C.}\ \bibnamefont {Liu}}, \bibinfo {author} {\bibfnamefont
  {R.-Y.}\ \bibnamefont {Liu}}, \bibinfo {author} {\bibfnamefont {R.-Z.}\
  \bibnamefont {Yang}}, \bibinfo {author} {\bibfnamefont {J.-Q.}\ \bibnamefont
  {Yu}}, \bibinfo {author} {\bibfnamefont {X.-G.}\ \bibnamefont {Ren}},
  \bibinfo {author} {\bibfnamefont {Z.-F.}\ \bibnamefont {Xu}}, \bibinfo
  {author} {\bibfnamefont {Y.-T.}\ \bibnamefont {Zhao}}, \bibinfo {author}
  {\bibfnamefont {Y.-S.}\ \bibnamefont {Huang}},\ and\ \bibinfo {author}
  {\bibfnamefont {J.-X.}\ \bibnamefont {Li}},\ }\bibfield  {title} {\bibinfo
  {title} {Signatures of linear {{Breit-Wheeler}} pair production in polarized
  {$\gamma\gamma$} collisions},\ }\href
  {https://doi.org/10.1103/PhysRevD.105.L071902} {\bibfield  {journal}
  {\bibinfo  {journal} {Phys. Rev. D}\ }\textbf {\bibinfo {volume} {105}},\
  \bibinfo {pages} {L071902} (\bibinfo {year} {2022})}\BibitemShut {NoStop}%
\bibitem [{\citenamefont {Han}\ \emph {et~al.}(2023)\citenamefont {Han},
  \citenamefont {Cai}, \citenamefont {Shou}, \citenamefont {Liu}, \citenamefont
  {Yu},\ and\ \citenamefont {Yan}}]{han2023Linear}%
  \BibitemOpen
  \bibfield  {author} {\bibinfo {author} {\bibfnamefont {L.~Q.}\ \bibnamefont
  {Han}}, \bibinfo {author} {\bibfnamefont {J.}~\bibnamefont {Cai}}, \bibinfo
  {author} {\bibfnamefont {Y.~R.}\ \bibnamefont {Shou}}, \bibinfo {author}
  {\bibfnamefont {X.~D.}\ \bibnamefont {Liu}}, \bibinfo {author} {\bibfnamefont
  {J.~Q.}\ \bibnamefont {Yu}},\ and\ \bibinfo {author} {\bibfnamefont {X.~Q.}\
  \bibnamefont {Yan}},\ }\bibfield  {title} {\bibinfo {title} {Linear
  {{Breit-Wheeler}} process driven by compact lasers},\ }\href
  {https://doi.org/10.1103/PhysRevE.108.055208} {\bibfield  {journal} {\bibinfo
   {journal} {Phys. Rev. E}\ }\textbf {\bibinfo {volume} {108}},\ \bibinfo
  {pages} {055208} (\bibinfo {year} {2023})}\BibitemShut {NoStop}%
\end{thebibliography}%

%\end{CJK}
\end{document}